\documentclass[reprint,superscriptaddress,amsmath,amssymb,aps,pra]{revtex4-2}
\usepackage{color}
\usepackage{booktabs}
\usepackage{makecell}
\usepackage{tabularx}
\usepackage{array}
\usepackage{graphicx}
\usepackage{subfigure}
\usepackage{stfloats}
\usepackage{caption}
\usepackage{multirow}
\usepackage{mathtools}
\usepackage{amsthm}
\usepackage{dcolumn}
\usepackage{bm}
\usepackage{algorithm}
\usepackage{algpseudocode}
\usepackage[colorlinks,linkcolor=blue,anchorcolor=blue,citecolor=blue]{hyperref}
\usepackage[T1]{fontenc}
\usepackage{orcidlink}

\begin{document}

\title{A Variational Approach to Unique Determinedness in Pure-state Tomography}

\author{Chao Zhang\,\orcidlink{0000-0002-2093-7496}}
\thanks{These authors contributed equally to this work}
\affiliation{Department of Physics, The Hong Kong University of Science and Technology, Clear Water Bay, Kowloon, Hong Kong, China}

\author{Xuanran Zhu\,\orcidlink{0000-0003-4754-9642}}
\thanks{These authors contributed equally to this work}
\affiliation{Department of Physics, The Hong Kong University of Science and Technology, Clear Water Bay, Kowloon, Hong Kong, China}

\author{Bei Zeng\,\orcidlink{0000-0003-3989-4948}}
\email{zengb@ust.hk}
\affiliation{Department of Physics, The Hong Kong University of Science and Technology, Clear Water Bay, Kowloon, Hong Kong, China}

\date{\today}

\begin{abstract}
In pure-state tomography, the concept of unique determinedness (UD)---the ability to uniquely determine pure states from measurement results---is crucial. This study presents a new variational approach to examining UD, offering a robust solution to the challenges associated with the construction and certification of UD measurement schemes. We put forward an effective algorithm that minimizes a specially defined loss function, enabling the differentiation between UD and non-UD measurement schemes. This leads to the discovery of numerous optimal pure-state Pauli measurement schemes across a variety of dimensions. Additionally, we discern an alignment between uniquely determined among pure states (UDP) and uniquely determined among all states (UDA) in qubit systems when utilizing Pauli measurements, underscoring its intrinsic robustness under pure-state recovery. We further interpret the physical meaning of our loss function, bolstered by a theoretical framework. Our study not only propels the understanding of UD in quantum state tomography forward, but also delivers valuable practical insights for experimental applications, highlighting the need for a balanced approach between mathematical optimality and experimental pragmatism.
\end{abstract}

\maketitle

\section{Introduction}
Quantum state tomography (QST) is a pivotal technique in quantum information science since it enables the accurate reconstruction and characterization of quantum states \cite{christandl2012reliable,lvovsky2009continuous,thew2002qudit,qi2013quantum,paris2004quantum}. As an essential tool in quantum devices and protocols, QST has far-reaching implications in various domains, including quantum computing \cite{steane1998quantum,o2007optical,preskill2018quantum}, quantum communication \cite{gisin2007quantum,cozzolino2019high,brassard2003quantum}, and quantum cryptography \cite{gisin2002quantum,pirandola2020advances,bennett1992quantum}.

General QST necessitates $d^2$ measurement outcomes to recover an arbitrary $d$-dimensional state. Several positive operator-valued measure (POVM) schemes, such as symmetric informationally complete POVM \cite{renes2004symmetric} and mutually unbiased bases POVM \cite{wootters1989optimal}, offer satisfactory state recovery. As for many-body systems, it has been demonstrated that a minimum of $3^n$ separable projective measurement settings are required for $n$-qubit systems \cite{de2008choice}. This number can be reduced to $2^n+1$ by allowing non-separable measurements \cite{adamson2010improving}. However, these measurement schemes for general QST can be prohibitively costly for experimental implementation due to the exponential complexity.

In quantum information science, the focus on pure states is driven by both theoretical inquiries, such as Pauli's foundational question about the uniqueness of wave functions determined by position and momentum distributions \cite{pauli1933allgemeinen}, and practical experimental applications, as evidenced by a variety of implementations \cite{xin2020quantum,ma2016pure,sosa2017experimental,liu2012experimental}. This interplay has spurred the development of pure-state tomography, a concept that not only stands on its own but also extends to rank-$r$ states \cite{heinosaari2013quantum,carmeli2014tasks,kech2017constrained,baldwin2016strictly} and matrix product states \cite{cramer2010efficient}.

The significance of pure-state tomography is threefold. Firstly, it offers resource efficiency by reducing the measurement requirements, owing to the known purity of the state. Secondly, accurate preparation and measurement of pure states are critical in many quantum information tasks like quantum computing and communication, making pure-state tomography indispensable for ensuring high-fidelity in these processes. Lastly, as a theoretical challenge, pure-state tomography deepens our understanding of many aspects of quantum mechanics such as quantum mariginal problem \cite{Klyachko_2006} and quantum state discrimination \cite{Bae_2015}.

In this study, we primarily focus on the problem about unique determinedness (UD) of pure states, given the specific measurement scheme $\mathbf{A}$ consists of observables $\{A_0=\mathbb{I},A_1, A_2,..., A_m\}$. A measurement scheme $\mathbf{A}$ is classified as UD if any pure state $|\psi\rangle$ is uniquely determined among pure states (UDP) or among all states (UDA) by measuring the given observables, i.e., any other pure state or mixed state cannot have the same measurement results as those of $|\psi\rangle$. These definitions are consistent with the notions of (strictly) informationally complete measurements \cite{flammia2005minimal,renes2004symmetric,scott2006tight}.

Necessary and sufficient conditions for UD measurement have been established through the examination of the eigenvalue structure of the orthogonal space with respect to $\mathbf{A}$ \cite{carmeli2014tasks,heinosaari2013quantum,PhysRevA.88.012109}. Interestingly, a gap between UDA and UDP has been identified in terms of the number of required observables \cite{heinosaari2013quantum,PhysRevA.88.012109}. A similar gap is also present for projective measurements \cite{carmeli2015many,carmeli2016stable}. In addition to the theoretical considerations, experimental aspects have been examined, such as the stability of state recovery. UDA has been proved to be more robust against the noise \cite{sosa2017experimental,gross2010quantum}.

However, constructing UD measurement schemes in a given scenario proves to be quite challenging, primarily due to the complexity of verifying UD's conditions about the eigenvalue structure. For example, in the study by Ma et al. \cite{ma2016pure}, optimal UDA Pauli measurement schemes were identified for 2- and 3-qubit systems. However, their approach relied on an exhaustive search, ultimately verifying the set by considering all possible linear combinations of the Pauli operators in the orthogonal space. This process led to lengthy and complex mathematical proofs, which are clearly not scalable or extendable to more general cases. To address this challenge, we propose an effective algorithm to determine whether a given measurement scheme is UD by minimizing a suitably defined loss function, which significantly streamlines the certification process.

In $n$-qubit systems, the numerical results show a clear gap in minimized loss between UD and non-UD measurement schemes. The former exhibits a discernibly nonzero value, while the latter approaches zero within the machine precision. Consequently, we can set a threshold $\delta$ based on the non-UD's minimum loss to determine whether a minimized loss is effectively zero or nonzero; if the value is above $\delta$, we can regard the scheme as UD.

Furthermore, with the assistance of random sampling techniques, we successfully identify numerous locally or globally optimal pure-state Pauli measurement schemes across various dimensions, including previous results for $2,3$-qubit UDA Pauli measurement \cite{ma2016pure}. Intriguingly, our findings reveal that in qubit systems, UDP invariably aligns with UDA when employing Pauli measurements, a phenomenon not commonly observed in other contexts. This insight implies that Pauli measurements inherently exhibit a convex property within the context of pure-state recovery.

However, our numerical analyses reveal that not all UD measurement schemes consistently exhibit distinct nonzero minimum losses as the dimension increases. A similar pattern is also noted in Pauli measurements, albeit not significantly. In essence, the gap between non-UD and UD schemes in terms of their minimum losses becomes increasingly less apparent, which poses the challenge for finding optimal schemes in higher dimension.  For instance, when considering the UD scheme built upon polynomial bases \cite{carmeli2015many,carmeli2016stable}, our numerical result shows such scheme experiences a marked decrease in minimum losses with increasing dimension. Through some theoretical analysis and numerical experiments, we find the relation between defined loss and the stability of state recovery. To elaborate, we observe that UD schemes with less minimum loss tend to exhibit greater instability in noisy state recovery, and also pinpoint those vulnerable states. These findings highlight a trade-off between the mathematical optimality and experimental pragmatism, as real-world experiments often further require noise-resilient UD schemes.

The structure of this paper is outlined as follows: In Sec.~\ref{pre}, we define some necessary concepts and summarize the previous results of UD schemes. In Sec.~\ref{method}, we introduce our variational approach for determining whether a given measurement scheme falls into the UD category. Furthermore, a search algorithm is proposed to identify the locally or globally optimal pure-state measurement scheme from discrete optional operators. In Sec.~\ref{results}, we investigate Pauli measurements that adhere to UD's criteria, unveiling several intriguing and unanticipated findings. In Sec.~\ref{analysis}, we explore the relationship between our defined loss function and the stability of pure-state tomography in the presence of noise, laying the groundwork for devising practical pure-state measurement schemes for real-world experiments. Finally, Sec.~\ref{discussions} offers additional discussions and insights.

\section{Preliminary}\label{pre}

Quantum state tomography has two basic ingredients: states and measurements. A quantum state in a $d$-dimensional Hilbert space $\mathcal{H}_d$ is denoted by a density matrix $\rho$, which is positive semi-definite and normalized to unit trace.

A measurement scheme corresponds to a set of observables $\mathbf{A}=\{A_0=\mathbb{I},A_1, A_2,..., A_m\}$, where the identity ensures the self-consistency of $\text{Tr}[\rho]=1$. A measurement process maps a quantum state $\rho$ to a real vector
\[ \mathbf{a}=\mathcal{M}_{\mathbf{A}}(\rho)=\left(\text{Tr}[A_0\rho], \text{Tr}[A_1\rho],...,\text{Tr}[A_m\rho]\right) \]
which we refer to as the measurement vector.

The kernel of a measurement scheme $\mathbf{A}$ is defined as the set of hermitian matrices with zero measurement vector
\[\text{Ker}(\mathbf{A})=\{x\in\mathbb{C}^{d\times d}:\mathcal{M}_{\mathbf{A}}(x)=\mathbf{0},x=x^\dagger\},\]
where we extend the domain of the mapping $\mathcal{M}_{\mathbf{A}}$ from quantum states to Hermitian matrices. Since the measurement set includes the identity, we immediately obtain that all elements in $\mathrm{Ker}(\mathbf{A})$ are traceless, i.e. $\forall x \in \mathrm{Ker}(\mathbf{A}),\text{Tr}[x]=0$. From the perspective of matrix space, $\mathrm{Ker}(\mathbf{A})$ is a Hermitian matrix space orthogonal to $\mathbf{A}$.

With the above concepts, we define UD measurement schemes as the following:

\textbf{Definition 1} (\textit{UDP scheme}). A measurement scheme $\textbf{A}$ is UDP if any pure state $|\psi\rangle$ can be uniquely determined among pure states by measuring $\textbf{A}$, i.e., there does not exist any other pure state which has the same measurement vector $\mathbf{a}$ as those of $|\psi\rangle$.

\textbf{Definition 2} (\textit{UDA scheme}). A measurement scheme $\textbf{A}$ is UDA if any pure state $|\psi\rangle$ can be uniquely determined among all states by measuring $\textbf{A}$, i.e., there does not exist any other state, pure or mixed, which has the same measurement vector $\mathbf{a}$ as those of $|\psi\rangle$.

In a $d$-dimensional quantum space, it has been established that a family of $4d-5$ observables suffices for a UDP scheme \cite{heinosaari2013quantum}, while for UDA the number is $5d-7$ \cite{PhysRevA.88.012109} (not including identity). Regarding $d$-dimensional POVMs ($d>2$), four orthonormal bases are deemed adequate for UDP \cite{carmeli2015many}. Nonetheless, for $d=4$, the sufficiency of three bases remains an unresolved question. In contrast, UDA demands five bases \cite{carmeli2016stable}.

Regarding the many-body systems, we have few previous results related to UD. One of them is the minimum number of Pauli operators for UDA in 2- and 3-qubit systems \cite{ma2016pure}. In the context of quantum compressed sensing, it is estimated that $O(d\log^2d)$ random Pauli measurements are adequate for UDA \cite{gross2010quantum}.

The research interest in UDA is not solely due to the fundamental distinction between UDP and UDA but also its experimental advantages. Previous studies have shown that a UDA measurement scheme is more resilient and efficient under the recovery process \cite{sosa2017experimental}, as it guarantees the convexity of the state recovery. This property is also implied in quantum compressed sensing \cite{gross2010quantum}, where low-rank state recovery using random Pauli measurements is stable against noise.

It has been shown that a necessary and sufficient condition for a measurement scheme to be UDP is that every nonzero $\Delta \in \text{Ker}(\mathbf{A})$ has $\max(n_{-},n_{+})\geq 2$ \cite{carmeli2014tasks}, where $n_{-}$ and $n_{+}$ are the number of strictly negative and positive eigenvalues of a matrix, respectively. Another equivalent statement is all nonzero elements $\Delta \in \text{Ker}(\mathbf{A})$ has $\text{rank}(\Delta)\geq 3$ \cite{heinosaari2013quantum}. For UDA, it is also proven that a necessary and sufficient condition is that every nonzero $\Delta \in \text{Ker}(\mathbf{A})$ has $\min(n_{-},n_{+})\geq 2$ \cite{carmeli2014tasks,PhysRevA.88.012109}. And another special but common case is that $\text{Ker}(\mathbf{A})=\{0\}$, i.e., measurement scheme $\mathbf{A}$ constructs a full tomography process. However, verifying these conditions related to eigenvalues presents considerable mathematical challenges. While there exist some elaborate mathematical proofs for specific scenarios, these proofs generally cannot be extended to a systematic approach.

\section{Method}\label{method}
Instead of using unextendible mathematical techniques, we introduce a variational approach to explore the specific eigenvalue structure of a certain matrix space, which makes the certification of a UD scheme much more efficient and automatic.

We can construct a variational matrix $\Delta(\vec{\lambda},\vec{\psi})$ to search a counterexample for the propositions about UDP/UDA. Starting with the simple one, UDP, we construct a Hermitian matrix with $n_{-}=1, n_{+}=1$, as the following:
\begin{equation}
   \Delta(\vec{\lambda},\vec{\psi}) = -\lambda_{1}|\psi_{1}\rangle\langle\psi_{1}|+\lambda_{2}|\psi_{2}\rangle\langle\psi_{2}|,\label{eq:udp-matrix-para}
\end{equation}
where $-\lambda_1,\lambda_2$ are one negative and one positive eigenvalues of $\Delta$ with the corresponding eigenvectors $|\psi_1\rangle,|\psi_2\rangle$ satisfying the orthonormal condition. In order to ensure the matrix $\Delta$ is nonzero, we fix the matrix Frobenius norm as $||\Delta||_\mathrm{F}^2=\lambda_1^2+\lambda_2^2=1$. In defining the loss function $\mathcal{L}$ as
\begin{equation}
	\mathcal{L}_{\mathbf{A}}(\vec{\lambda},\vec{\psi})=\left\Vert \mathcal{M}_{\mathbf{A}}\left(\Delta(\vec{\lambda},\vec{\psi})\right)\right\Vert _{2}^{2}. \label{eq:loss-function}
\end{equation}
Here, we utilize the square of norm-2 for its computational simplicity and smoother gradient properties, avoiding the potential instability in optimization that norm-1's nonzero gradient near zero might cause. If the loss function cannot be minimized to zero (up to the machine precision), in principle, we can conclude that all nonzero elements $\Delta \in \text{Ker}(\mathbf{A})$ has $\max(n_{-},n_{+})\geq 2$, i.e., $\textbf{A}$ belongs to UDP.

Similarly, for UDA, we can construct a variational matrix $\Delta$ with one negative eigenvalue and at most $d-1$ positive eigenvalues, i.e., $n_{-}=1,n_{+} \leq d-1$:
\begin{equation}
	\Delta(\vec{\lambda},\vec{\psi}) = -\lambda_{1}|\psi_{1}\rangle\langle\psi_{1}|+\sum_{i=2}^{d}\lambda_{i}|\psi_{i}\rangle\langle\psi_{i}|.\label{eq:uda-matrix-para}
\end{equation}
The orthogonal and normalized eigenvectors $|\psi_i\rangle$ are generated from a unitary matrix via matrix exponential

\[f({\theta_\psi})=\mathrm{exp}({\theta_\psi}-\theta_\psi^T+i{\theta_\psi}+i\theta_\psi^T):\mathbb{R}^{d\times d}\to\mathbb{C}^{d\times d} ,\]
where the $i$-th column of $f(\theta_{\psi})$ corresponds to $|\psi_i\rangle$.
Again, we require the matrix $\Delta$ to be normalized $\Vert\Delta\Vert_{\text{F}}^2=\sum_i\lambda_i^2=1$. To generate those non-negative valued $\lambda_i$, we use the Softplus function
\[\text{Softplus}(\theta_\lambda)=\log(1+e^{\theta_\lambda}):\mathbb{R}\to\mathbb{R}^+,\]
which is quite common in the machine learning area \cite{softplus}. Softplus function serves as a smooth approximation to the ReLU function \cite{relu} and is employed to ensure that the output is invariably positive. Meanwhile, the output of the Softplus can approach zero up to machine precision, therefore the variational matrix $\Delta$ in (\ref{eq:uda-matrix-para}) can effectively represent the matrix with $n_+\leq d-1$ from the numerical perspective. The same loss function $\mathcal{L}$ in (\ref{eq:loss-function}) can be used for the UDA case. Moreover, since the matrix $-\Delta$ models the element with eigenvalues $n_+=1,n_-\leq d-1$ and gives the same loss function, there is no need to include $-\Delta$ in the discussion explicitly. With the same argument, we can say a measurement scheme $\mathbf{A}$ is UDA if we cannot optimize the loss function $\mathcal{L}$ to zero within the limits of machine precision.

To minimize the loss function $\mathcal{L}$, we use gradient-based optimization, updating parameters $\theta_\psi$ and $\theta_\lambda$ via PyTorch's gradient back-propagation \cite{pytorch}. Unlike typical neural networks, our loss function lacks inherent randomness, leading us to choose the L-BFGS-B method from the SciPy package \cite{2020SciPy-NMeth}, which generally converges more rapidly than standard neural network training algorithms like SGD or Adam. To mitigate the risk of converging only to local minima, we implement a probabilistic approach: performing $N$ trials with initially normally distributed $\theta_\lambda$ and $\theta_\psi$, followed by gradient descent updates. A higher number of trials ($N$) improves the chances of reaching the global minimum, though at a cost of increased computation time. The choice of $N$ should be adjusted based on the desired balance between computational resources and the risk of settling at local minima.

However, the question arises: how can we ascertain whether a minimized loss is effectively zero or nonzero? To address this, we introduce another hyper-parameter, the threshold $\delta$. This threshold is determined based on the minimum loss for non-UD cases, which is considered as zero. By comparing the minimized loss with this threshold, we can discern whether the obtained minimum loss is effectively zero or nonzero.

For instance, one UDA Pauli measurement with the minimum size for 2-qubit has the following form \cite{ma2016pure}:
\[ \mathbf{A}=  \{I I, I X, I Y, I Z, X I, Y X, Y Y, Y Z, Z X, Z Y, Z Z\}, \]
which leads to a minimum loss of $1.0$. Notably, if an operator is removed, the corresponding minimized loss will immediately deteriorate to the level of $10^{-11}$. Thus, $10^{-11}$ can be a reference for setting the threshold $\delta$.

We then propose an algorithm based on random sampling to search the locally or even globally optimal UDA measurement schemes from discrete optional operators, as shown in Algorithm \ref{algo:uda}, and the UDP one will be similar except replacing the matrix $\Delta$ from (\ref{eq:uda-matrix-para}) with (\ref{eq:udp-matrix-para}). Here, the locally (globally) optimal means the size of the operator set is a local (global) minimum and removing any operator from the algorithm output will lead to a non-UDA scheme.

The algorithm takes inputs: a set of optional observables $\textbf{A}$ and some hyper-parameters (number of trials $N$ and threshold $\delta$). The variable $F$ is a set containing the operators that must be included in the measurement scheme, which is initialized with the identity $I$. At each step, the algorithm strategically removes an element from the set $\textbf{A}$ and checks if the resulting observable set still meets the UDA criteria. If the kernel $\mathrm{Ker}(\textbf{A})$, consists solely of the zero matrix, we can bypass the more time-intensive optimization steps, as this scenario unequivocally confirms UDA. For the input set $\textbf{A}$ of finite size $n$, the algorithm's for-loop will iterate a maximum of $n$ times. In the end, such an algorithm will output a locally or globally optimal measurement scheme.

\begin{algorithm}[H]
\caption{UDA: Input a set of observables $\textbf{A}$, number of trials $N$ and threshold $\delta$. Return a measurement set $A$ with the minimum size}\label{algo:uda}
\begin{algorithmic}[1]
\Procedure{T}{$\textbf{A}$}\Comment{criteria for UDA}
    \If {$\text{Ker}(\textbf{A})=\{0\}$ or $\min_\Delta\mathcal{L}_{\textbf{A}}(\Delta)>\delta$ for $N$ trials}
        \State \textbf{return} True
    \Else
        \State \textbf{return} False
    \EndIf
\EndProcedure
\Procedure{UDA}{$\textbf{A},N,\delta$}
    \State $F\leftarrow \left\{I\right\}$ \Comment{init with identity matrix}
    \While{$\textbf{A}\ne F$}
        \State randomly pick $x$ from $\textbf{A}\backslash F$
        \If {$T(\textbf{A}\backslash \left\{x\right\})$}
            \State $\textbf{A}\gets \textbf{A} \backslash \left\{x\right\}$
        \Else
            \State $F\gets F\cup \left\{x\right\}$
        \EndIf
    \EndWhile
    \State \textbf{return} $\textbf{A}$
\EndProcedure
\end{algorithmic}
\end{algorithm}

\section{Results}\label{results}

In this section, we focus on Pauli measurement, i.e., $\textbf{A} \subseteq \{I,X,Y,Z\}^{\otimes n}$.
A natural inquiry emerges: what is the minimum number of Pauli operators required to realize pure-state tomography in qubit systems? This question gives rise to a combinatorial optimization problem, which can be highly intricate. While results for 2- and 3-qubit systems have been derived using unextendible mathematical proofs \cite{ma2016pure}, the general case remains undetermined.

Nevertheless, our method enables the identification of locally and even globally optimal Pauli measurement schemes for pure-state tomography. It is noteworthy that if a set of Pauli operators is classified as UD, any set that is unitarily equivalent to it is inherently UD. A particular class of unitary operators which maps the set of Pauli operators to itself is called Clifford group. The order of that group increases substantially with the number of qubits; for instance, 1-, 2-, and 3-qubit systems correspond to 24, 11520, and 92897280 elements, respectively. This observation implies the diversity of pure-state tomography schemes in Pauli measurement, as suggested in quantum compressed sensing \cite{gross2010quantum}. Furthermore, numerical experiments reveal that the minimum loss values between UD and non-UD Pauli measurements exhibit a clear gap across the variety of dimensions, implying a relatively large threshold $\delta$ is sufficient.

\begin{figure}[H]
    \centering
    \includegraphics[width=\linewidth]{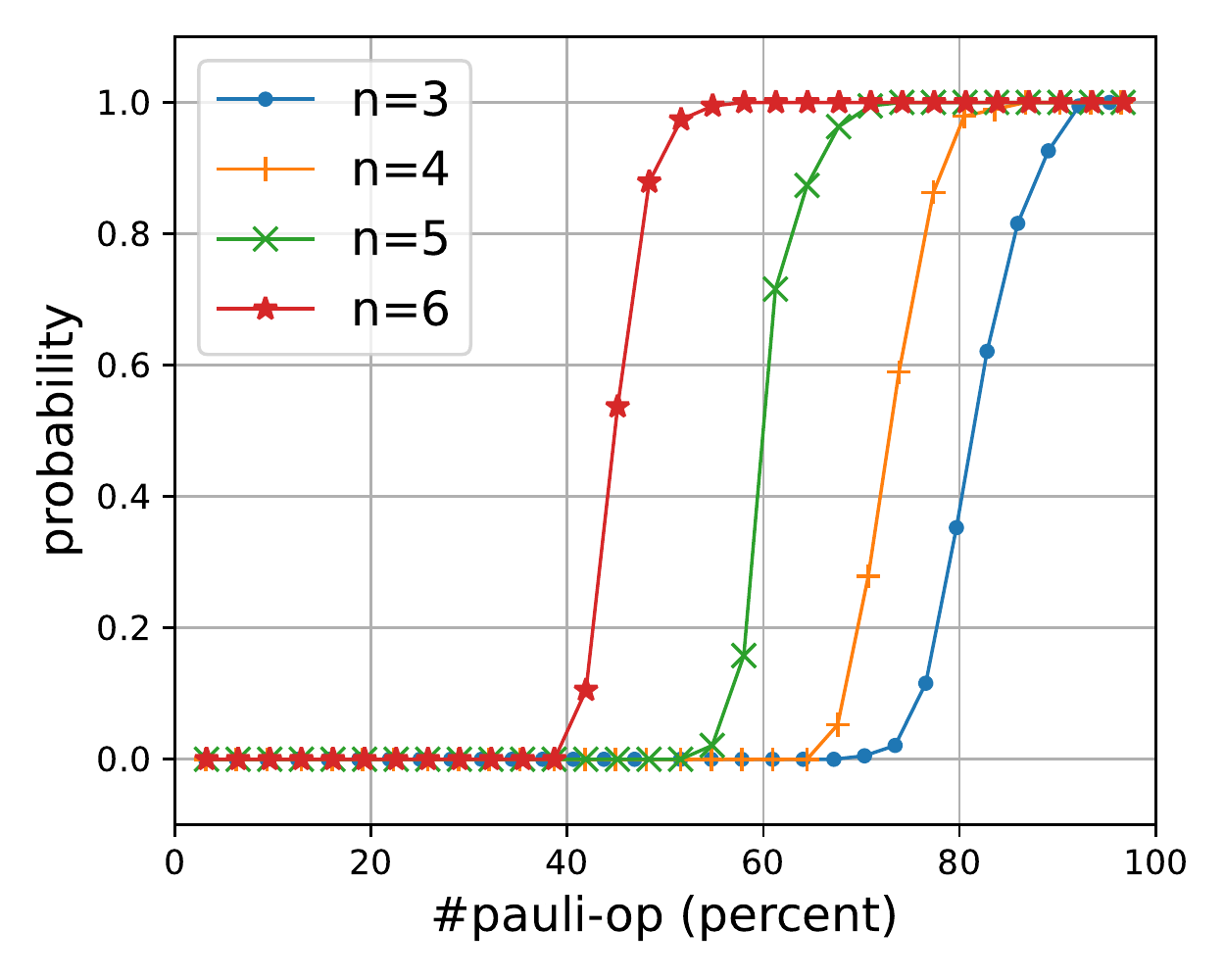}
    \caption{The graph depicts the transition between low-probability and high-probability regions for varying percentages of operators in the whole Pauli group. For each data point, we generate 190 random subsets and employ the hyperparameters $\delta=0.01$ and $N=80$ in our algorithm to determine whether a subset qualifies as a UDA measurement scheme. Subsequently, we calculate the corresponding qualified probability. }
    \label{fig:n-qubits-pauli}
\end{figure}

We first revisit the properties for randomly sampled Pauli operator subsets, and the results are shown in FIG. \ref{fig:n-qubits-pauli}. From $n=3$ qubits systems to $n=6$ qubits systems, Pauli subsets with different sizes are randomly sampled and then the probabilities of being UDA are computed. As the number of Pauli operators increases, there is a quick shift between low-probability and high-probability regions for UDA. This result aligns with the compressed sensing framework presented in \cite{gross2010quantum}, which demonstrates that randomly sampled Pauli operators with a certain number can be utilized effectively for low-rank state tomography in the presence of noise.

We then apply Algorithm \ref{algo:uda} to search locally or even globally optimal Pauli operator sets satisfying the UD criteria. Firstly, an interesting result found by the algorithm is that all UDP schemes are also UDA. This phenomenon is not commonly observed in other contexts, implying the Pauli measurement inherently exhibits a convex property within the realm of pure-state tomography. Making use of this phenomenon, we can run the algorithm to search for optimal UDP Pauli schemes and then verify whether the found one is a UDA scheme since the UDA searching algorithm is much more time-consuming than the UDP one. Due to that alignment, we will uniformly denote them as UD. Secondly, deduced from our random sampling experiment (see FIG. \ref{fig:n-qubits-pauli}), let $k_n$ indicates a threshold for the initial random subset size of Pauli operators. For a system of $n$ qubits with $4^n$ Pauli operators, a subset of more than $k_n$ operators typically constitutes a UD measurement with high probability. This approach enables us to start with a smaller subset rather than the entire Pauli group, enhancing computational efficiency. If this subset fails the UD criteria, we repeat the sampling process. For 3, 4, and 5-qubit systems, we selected initial subsets of approximately $0.85/0.7/0.55 \times 4^n \approx 54/176/554$ operators, respectively, as observed to be effective in our numerical experiments.

The outcomes of the search algorithm are presented in TABLE \ref{tab:pauli-ud}. Through the application of this algorithm, a multitude of locally or globally optimal UD schemes have been discovered, with several of them listed in the table. For 2- and 3-qubit systems, we have successfully identified the optimal UD schemes with minimal size, aligning with previous findings reported in \cite{ma2016pure}. For the 4-qubit and 5-qubit Pauli measurements, the minimal size of the UD schemes found is $106$ and $393$ respectively, which are findings previously unreported in the literature. Upon a thorough investigation of all Clifford elements \cite{Koenig_2014}, we have ascertained that within the 2-qubit system, all 6 discovered subsets, each comprising 11 Pauli operators, are equivalent. Additionally, there are 20 equivalent subsets each containing 13 operators, of which we have successfully identified 19. In the context of a 3-qubit system, our findings show that there are precisely 30,240 equivalent subsets, each with 31 Pauli operators; our research has uncovered 176 of these subsets. As we extend our analysis to the 4-qubit scenario, the number of equivalent subsets, each encompassing 106 operators, surpasses 100 million at least.

\begin{table}[H]
    \centering
    \caption{\label{tab:pauli-ud}UD scheme with Pauli measurement. The column $m\times n$ denotes we find $m$ UD Pauli measurement schemes with $n$ Pauli operators (including the identity), which could be Clifford equivalent. The minimized loss $\mathcal{L}$ for UDA and UDP is evaluated for the scheme with minimum size. The computation time (in second) is also listed, wherein $t_1$ and $t_2$ respectively represent the time required for a single certification and the total time necessitated to search for a locally optimal scheme. $N$ denotes the number of trials in a single certification.}
    \begin{tabular}{ccccc}
        \toprule
        \#qubits & (UDA,UDP) $\mathcal{L}$ & $m\times n$ & $t_1/t_2$ & $N$\\ 
        \midrule
        $2$ & $(1,2)$ & $6\times11,19\times 13$ & $0.05/0.4$ & $10$\\
        $3$ & $(0.519,2)$ & $176\times 31,258\times 32$ & $0.06/2.5$ & $10$\\
        $4$ & $(0.280,1.788)$ & $3\times 106,14\times 107$ & $2.4/81$ & $80$\\
        $5$ & $(0.202,1.951)$ & $1\times 393,1\times 395$ & $25/8400$ & $640$\\
        \bottomrule
    \end{tabular}
\end{table}

Consideration is also given to the factor of time scaling. As observed in the table, the computation time, denoted as $t_2$, required for locating a locally optimal scheme escalates significantly with an increase in the number of qubits. This escalation imposes constraints on feasibility in higher dimensions. However, in practical applications, a randomly sampled subset of Pauli operators can already dramatically mitigate the number of required operators. The time needed to identify such a subset equates to a relatively small integer multiple of $t_1$, rendering it acceptable in most circumstances. 
\section{Analysis}\label{analysis}

The numerical results demonstrate the feasibility of our method for identifying optimal pure-state Pauli measurement schemes. However, we observe that as the dimension increases, not all types of UD measurement schemes consistently exhibit a distinct nonzero minimum loss. This presents a challenge for the discovery in higher dimensions.  Indeed, the case for Pauli measurements also decays, albeit marginally. One might speculate that increasing the fixed norm along with the dimension could improve the situation, but this would affect not only the minimum loss for UD but also that of non-UD, which remains the difficulty of determining the threshold $\delta$.
The above observation prompts us to consider the physical meaning of our previously defined loss function (\ref{eq:loss-function}), and finally reveals a connection to the stability against noise. This connection implies a trade-off between the mathematical optimality and experimental pragmatism. 

\subsection{Stability theory}
A realistic measurement is always affected by noise, which leads to the deviation of the measurement results. For that reason, here, we define the closeness of two measurement schemes. Two measurement schemes $\mathcal{M}_{\mathbf{A}},\mathcal{M}_{\mathbf{A^{\prime}}}$ are $\epsilon$-closed if
\[ \max_{x=x^\dagger,\lVert x\rVert_{\text{F}=1}} \lVert\mathcal{M}_{\mathbf{A}}(x)-\mathcal{M}_{\mathbf{A'}}(x)\rVert_2 <\epsilon.\]
Then two theorems can be obtained directly from \cite{carmeli2016stable} with different norms.

\textbf{Theorem 1} (\textit{Stability of unique determinedness}). If a measurement scheme \textbf{A} is UD, then there is an $\epsilon>0$ such that every measurement scheme $\textbf{A}^{\prime}$ which is $\epsilon$-closed to \textbf{A} is also UD.

\begin{proof}Let
\[ K_1 := \{x\in \mathbb{C}^{d\times d}: x=x^\dagger,\lVert x\rVert_{\text{F}}=1,n_-(x)=n_+(x)=1\} \]
and
\[ K_2 := \{x\in \mathbb{C}^{d\times d}: x=x^\dagger,\lVert x\rVert_{\text{F}}=1,n_{-}(x)=1\}.\]
As we have seen, a measurement scheme is UD if and only if
\[ c:=\min_{x\in K_s}||\mathcal{M}_{\textbf{A}}(x)||_2>0 \]
with $s=1$ for UDP and $s=2$ for UDA. Since $\mathcal{M}_{A'}$ is $\epsilon$-closed with respect to $\mathcal{M}_A$, we have
\begin{align*}
    &\min _{x \in K_s}\lVert\mathcal{M}_{\textbf{A}'}(x)\rVert_2\\
    \geq& \min_{x\in K_s} \left( \lVert \mathcal{M}_{\textbf{A}}(x)\rVert_2 - \lVert \mathcal{M}_{\textbf{A}}(x)-\mathcal{M}_{\textbf{A}'}(x) \rVert_2  \right)\\
    \geq& c-\epsilon
\end{align*}
Hence, for any $\epsilon<c$, the measurement scheme $\textbf{A}^{\prime}$ is also UD.
\end{proof}

Clearly, we can find that the minimum value of the loss function $\mathcal{L}_{\textbf{A}}$ will reflect the stability of a UD measurement scheme \textbf{A}. The allowed systematic error for pure-state measurement is bounded by $c$, which is the square root of our loss function's minimum value. Furthermore, we can consider the state recovery process:
\begin{equation}
\begin{gathered}
\operatorname{minimize}\quad \left\Vert \mathcal{M}_{\mathbf{A}}(Y)-b\right\Vert _{2} \\
\text { s.t. } \begin{cases}
Y\succeq0, \text{Tr}[Y]=1\\
\mathrm{rank}(Y)=1 & \mathrm{(UDP\;only)}
\end{cases}\\
\end{gathered}\label{eq:SDP}
\end{equation}
where $b:=\mathcal{M}_{\textbf{A}}(\sigma)+f$ is the perturbed measurement data with a noisy term $f$. It should be emphasized that the UDP version is not a convex optimization problem and does not guarantee to find the global minimizer $Y^*$ using gradient-based methods while UDA does. For the noiseless case $\|f\|_2=0$, the minimizer $Y^*$ will be exactly the same as the original state $\sigma$. For some nonzero noise $\|f\|_2>0$, we define the stability coefficient $\alpha$ as the ratio of the state recovery error over the summation of the optimization objective function and the noise rate
\[ \alpha:=\frac{\left\|Y^*-\sigma\right\|_{\text{F}}}{\left\Vert \mathcal{M}_{\mathbf{A}}(Y^*)-b\right\Vert _{2}+\|f\|_2}. \]
The Frobenius norm is chosen for its convenience in numerical calculations. In order to get an accurate restored state, the stability coefficient $\alpha$ is expected to be as small as possible. Then the second theorem arises:

\textbf{Theorem 2} (\textit{Stability of recovery}). Given a UD measurement scheme $\textbf{A}$, for all pure state $\sigma$ and error term $f$, the stability coefficient $\alpha$ is bounded
\begin{equation}
    \alpha\leq \left( \inf_{x\in K_s} \left\Vert \mathcal{M}_{\mathbf{A}}(x)\right\Vert _{2}\right)^{-1} \label{eq:alpha-bound}
\end{equation}
with $s=1$ for UDP and $s=2$ for UDA.

\begin{proof}
We will prove the UDA case in the following since UDP can be done in the same way. Let $Y^*$ be the minimizer of the optimization problem (\ref{eq:SDP}). From the definition of the perturbed measurement $b$ and the Cauchy-Schwarz inequality, we have
$$
\begin{aligned}
&\left\|\mathcal{M}_{\textbf{A}}\left(Y^*\right)-b\right\|_2\\
\geq&\left\|\mathcal{M}_{\textbf{A}}\left(Y^*-\sigma\right)\right\|_2-\|f\|_2 \\
\geq&\left\|Y^*-\sigma\right\|_{\text{F}} \inf _{\substack{X  \succeq 0, X \neq \sigma\\ \text{Tr}[X]=1}} \left\|\mathcal{M}_{\textbf{A}}\left(\frac{X-\sigma}{||X-\sigma||_{\text{F}}}\right)\right\|_2-\|f\|_2 \\
\geq & ||Y^{*}-\sigma||_{\text{F}}\cdot \inf_{x \in K_2}||\mathcal{M}_{\textbf{A}}(x)||_2-\|f\|_2
\end{aligned}.
$$
The fourth line is derived from the following inclusion relation derived from Weyl's inequalities.
\[ \left\{\frac{X-\sigma}{\left\|X-\sigma\right\|_\text{F}}: X \succeq 0, X \neq \sigma,\\ \text{Tr}[X]=1\right\} \subseteq K_2. \]
\end{proof}

Consequently, the upper bound of the coefficient $\alpha$ exhibits an inverse relationship with $\min_{x \in K_s}||\mathcal{M}_{\textbf{A}}(x)||_2$, which corresponds to the square root of the minimal value of the loss function. This quantity can be regarded as a quality factor, signifying the effect of noise in the state recovery.

In summary, UDA schemes offer enhanced reliability in recovery due to their convexity, while UDP schemes may encounter local minima, resulting in the inability to obtain the optimal minimizer $Y^{*}$. On the other hand, measurement schemes with lower minimum loss demand greater experimental precision and yield suboptimal recovery outcomes.

We demonstrate the theorems above in the following example: a projective measurement scheme constructed by polynomial bases, which highlights the significance of minimum loss in pure-state tomography against noise.

\subsection{Measurement with polynomial bases}
\label{polynomial}
The four polynomial bases (4PB) \cite{carmeli2015many} for one qudit $\mathcal{H}_d$ consists of the following 4 orthonormals
\begin{gather*}
   \mathcal{B}_{1}=\big\{\left|\phi_{k}\right\rangle \left\langle \phi_{k}\right|:\left|\phi_{k}\right\rangle =\sum_{j=0}^{d-1}p_{j}\left(x_{d,k}\right)\left|j\right\rangle \big\}_{k=1}^{d}\\
   \mathcal{B}_{2}=\big\{\left|\phi_{k}\right\rangle \left\langle \phi_{k}\right|:\left|\phi_{k}\right\rangle =\sum_{j=0}^{d-2}p_{j}\left(x_{d-1,k}\right)\left|j\right\rangle \big\}_{k=1}^{d-1}\cup\left\{ P_{d-1}\right\}\\
   \mathcal{B}_{3}=\big\{\left|\phi_{k}\right\rangle \left\langle \phi_{k}\right|:\left|\phi_{k}\right\rangle =\sum_{j=0}^{d-1}e^{ij\alpha}p_{j}\left(x_{d,k}\right)\left|j\right\rangle \big\}_{k=1}^{d}\\
   \mathcal{B}_{4}=\big\{\left|\phi_{k}\right\rangle \left\langle \phi_{k}\right|:
   \left|\phi_{k}\right\rangle =\sum_{j=0}^{d-2}e^{ij\alpha}p_{j}\left(x_{d-1,k}\right)\left|j\right\rangle \big\}_{k=1}^{d-1}\cup\left\{ P_{d-1}\right\}
\end{gather*}
where $p_j(x)$ can be any normalized orthogonal polynomials, $x_{d,1},x_{d,2},\cdots,x_{d,d}$ are roots of the $d$-th order polynomials $p_d(x)$, and $P_i=\left|i\right\rangle \left\langle i\right|$ is the projector to the $i$-th basis. Also, the condition $e^{ij\alpha}\notin\mathbb{R},j=1,2,\cdots,d-1$ is required. Combining $\mathcal{B}_{5}=\big\{ P_{i}\big\}_{i=0}^{d-1}$ with the 4PB, we can obtain 5PB. Below we choose $p_n(x)$ as Chebyshev polynomials, $\alpha=\frac{\pi}{d}$ for simplicity.

\begin{figure}[H]
    \centering
    \includegraphics[width=\linewidth]{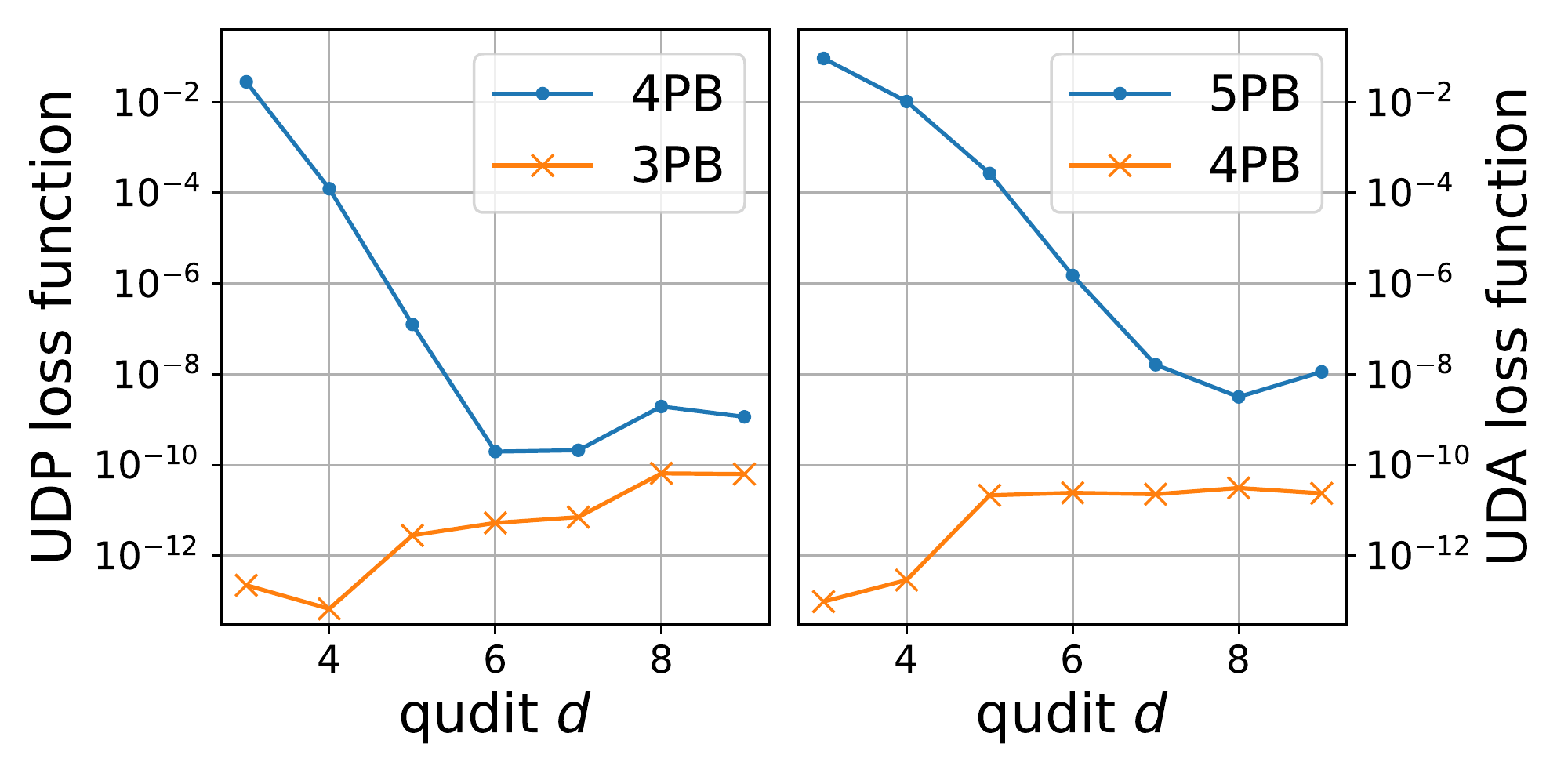}
    \caption{\label{fig:3PB-4PB-5PB}(a) the UDP loss function comparing 4PB and 3PB (drop $\mathcal{B}_4$ from 4PB). (b) the UDA loss function comparing 5PB and 4PB. The blue dotted lines denote the UDP/UDA schemes, while the orange crossed lines denote the non-UDP/UDA schemes. The observed fluctuations in the minimum losses below $10^{-8}$ can be attributed to limitations in convergence tolerance.}
\end{figure}

These projectors compose our measurement scheme $\textbf{A}$, where we omit the identity operator for convenience. It has been proven that 4PB is UDP \cite{carmeli2015many}, while 5PB is UDA \cite{carmeli2016stable} for any $d$-dimesional quantum space. The absence of any base will result in non-UD.  We calculate the minimum losses for different dimensions between UDP/UDA and non-UDP/UDA in FIG. \ref{fig:3PB-4PB-5PB}. Notably, the minimum losses for non-UD consistently remain below $10^{-10}$, while UD's minimum losses tend to approach zero as the dimension increases, thereby introducing the potential for instability.

\begin{figure}[H]
    \centering
    \includegraphics[width=\linewidth]{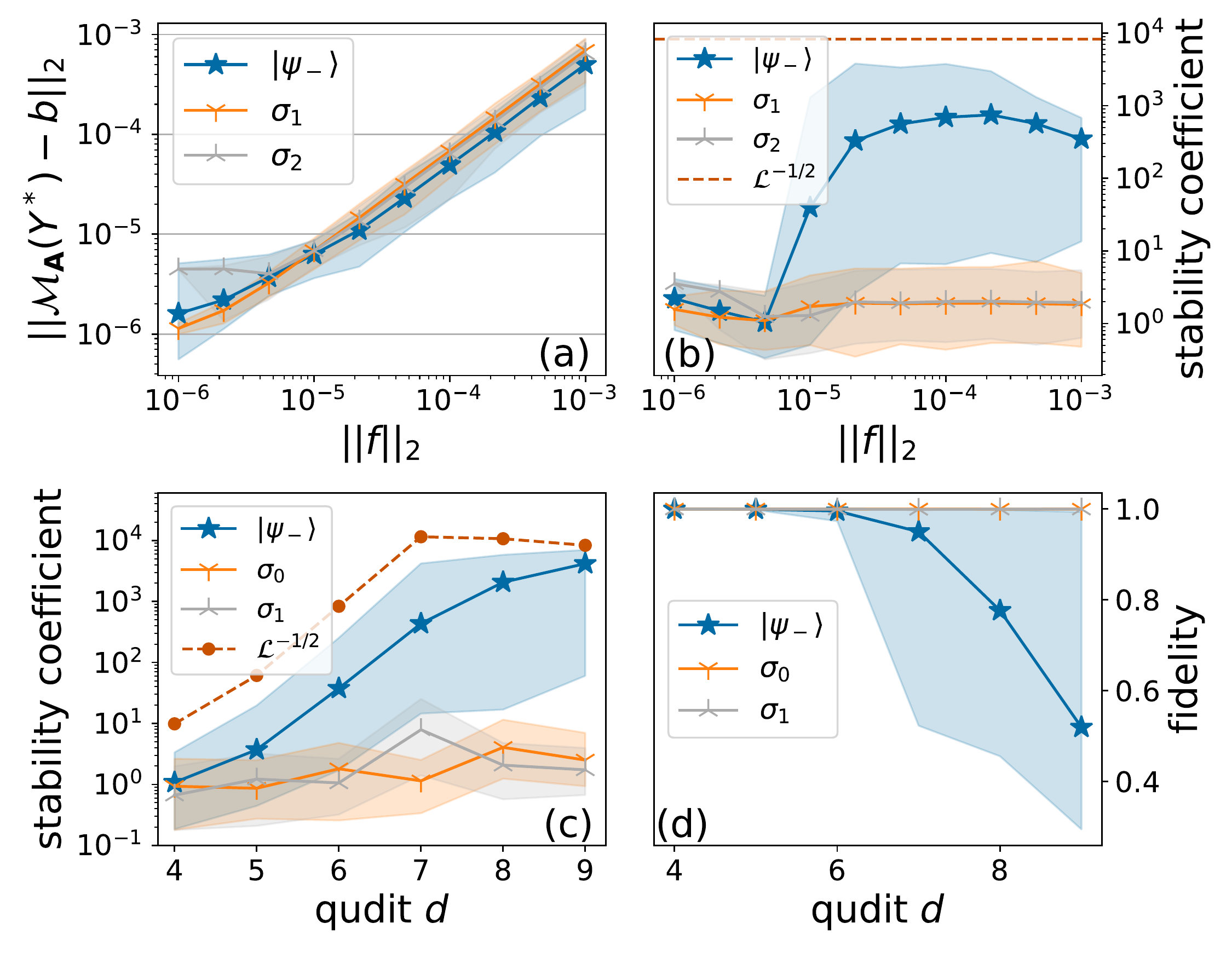}
    \caption{Stability of state recovery using 5PB (UDA) for one $d$-dimensional qudit. $|\psi_-\rangle$: the pure state with the negative eigenvalue found in minimizing the loss function (\ref{eq:loss-function}), and $\sigma_1,\sigma_2$: two random pure states. (a) $d=7$, the recovered error $\lVert\mathcal{M}_A(Y^*)-b\rVert_2$ versus noise rate $\lVert f\rVert_2$. (b) stability coefficient $\alpha$ versus noise rate $\lVert f\rVert_2$. (c) stability coefficient $\alpha$ versus dimension $d$. (d) fidelity between the original and the restored states for different dimensions $d$. Shaded region denotes (minimum, maximum) range over randomly sampled noise for all subfigures.}
    \label{fig:5PB-worst-case}
\end{figure}

We then investigate the stability of state recovery of the 5PB, and the results are shown in FIG. \ref{fig:5PB-worst-case}. Given the measurement scheme 5PB, we can find the optimal $\Delta^*$ in (\ref{eq:uda-matrix-para}) with $1$ negative eigenvalue and $d-1$ positive eigenvalues to give the minimum of the loss function in (\ref{eq:loss-function}). We conjecture that the eigenvector $|\psi_-\rangle$ with the negative eigenvalue will produce the worst performance of state recovery, according to the upper bound proposed in (\ref{eq:alpha-bound}). To verify this conjecture, we generate two random pure states $\sigma_0,\sigma_1$ according to Haar measure \cite{haar-measure} for comparison.

In sub-figure (\ref{fig:5PB-worst-case}a), we fix qudit dimension $d=7$ and randomly generate measurement results $b=\mathcal{M}_A(\sigma)+f$ with different noise rate $\lVert f\rVert_2$. $500$ random noise vectors $f$ are generated for every noise rate, then convex optimization package cvxpy \cite{diamond2016cvxpy} is used to solve this semi-definite programming in (\ref{eq:SDP}) and the minimizer $Y^*$ is always found. The average recovery errors $\lVert \mathcal{M}_A(Y^*)-b\rVert_2$ over $500$ samples are plotted in solid lines with the shaded region denoting the minimum-maximum range. From the figure, we can see that the recovery error in measurement space is almost linear with respect to the noise rate for all three pure states.

However, the recovery errors $\lVert Y^*-\sigma\rVert_\text{F}$ in density matrix space present a significant difference for $|\psi_-\rangle$ state as shown in sub-figure (\ref{fig:5PB-worst-case}b). When the noise rate is relatively low (e.g. $\lVert f\rVert_2\approx 10^{-6}$), the stability coefficient $\alpha$ are all below $10$ which is acceptable. In other words, the restored state $Y^*$ is pretty close to the original state $\sigma$. For a not-too-large noise rate (e.g. $\lVert f\rVert_2\approx 2\times 10^{-5}$), the stability coefficients for two random states remain below $10$, while the coefficient for $|\psi_-\rangle$ explodes to around $1000$. As a reference, we also plot the predicted upper bound (\ref{eq:alpha-bound}) in red dashed line $\mathcal{L}^{-1/2}=8260$. A clear transition can be observed around $\lVert f\rVert_2=10^{-5}$.

In sub-figure (\ref{fig:5PB-worst-case}c), for a fixed noise rate $\lVert f\rVert_2=10^{-4}$, the stability behavior versus different qudit dimension $d$ is investigated. We find that such instability is getting worse as the dimension $d$ increases. Still, the upper bound proposed in (\ref{eq:alpha-bound}) is shown in a red dashed line for reference. It should be emphasized that the upper bound $\mathcal{L}^{-1/2}$ will be larger than $10^4$ for $d\geq 8$, and the deceptive saturation in the figure is due to the limited precision in the numerical calculations. Subsequently, we compute the corresponding fidelities as illustrated in sub-figure (\ref{fig:5PB-worst-case}d). It becomes evident that the fidelity for the worst-case scenario diminishes considerably as the dimension $d$ increases. For comparison, we also explore the stability of pure-state tomography employing Pauli measurement (see Appendix \ref{pauli}), which gives almost perfect state recovery.

In conclusion, generally, for random states (like $\sigma_0,\sigma_1$ in the figure), the restored states will be close to the original ones. But there are some special states (like $|\psi_-\rangle$ found in optimization), once the noise rate is greater than some threshold, a not-too-large noise can lead to a significantly different restored state, which implies pursuing strictly optimal pure-state tomography with minimal loss might not be a practical approach when considering the stability against noise. 

\section{Discussions}\label{discussions}
In this study, we tackle the challenge of verifying UD measurement schemes by proposing an efficient algorithm that minimizes a suitably defined loss function. This method allows us to determine if a given measurement scheme is qualified for pure-state tomography, which can also be applicable for measurement settings \cite{li2017optimal}

Using random sampling techniques, we identify numerous optimal pure-state Pauli measurement schemes across various dimensions and discover that in qubit systems, UDP always coincides with UDA for Pauli measurements, suggesting a convex property in pure-state tomography. However, the mathematical proof of this property needs to be further investigated.

Our findings also indicate that not all types of UD schemes exhibit a obviously nonzero value as the dimension increases, making the distinction between non-UD and UD schemes less evident. By exploring the connection between our loss function and the stability of measurement schemes, we reveal that pure-state measurement scheme with a lower minimum loss will result in suboptimal state recovery. We investigate and demonstrate this phenomenon in the projective measurement constructed by polynomial bases. These findings suggest a trade-off between the mathematical optimality of measurement schemes and their stability against noise.

The utility of our proposed methodology extends beyond the realm of pure-state tomography, encompassing additional tasks in the domain of quantum information, such as entanglement detection \cite{G_hne_2009} and super-activation \cite{duan2009super}. Our approach has the potential to evolve into a comprehensive framework for scrutinizing the unique structure of a given space, which could comprise desired elements such as operators, matrices, and states.

For instance, with slight modifications, our method can be used to detect entanglement or even higher Schmidt ranks within a given subspace. This approach could prove to be more efficient and universal compared to recent work \cite{johnston2022complete}. Detailed elaboration and exploration of these potential applications will be presented in our forthcoming work.

\acknowledgments

We gratefully acknowledge Jianxin Chen and Jie Zhou for insightful discussions and the assistance of ChatGPT in facilitating the writing process. C. Zhang, X-R. Zhu and B. Zeng are supported by GRF grant No. 16300220.
\vspace{0.8cm}

\begin{figure}[H]
    \centering
    \includegraphics[width=\linewidth]{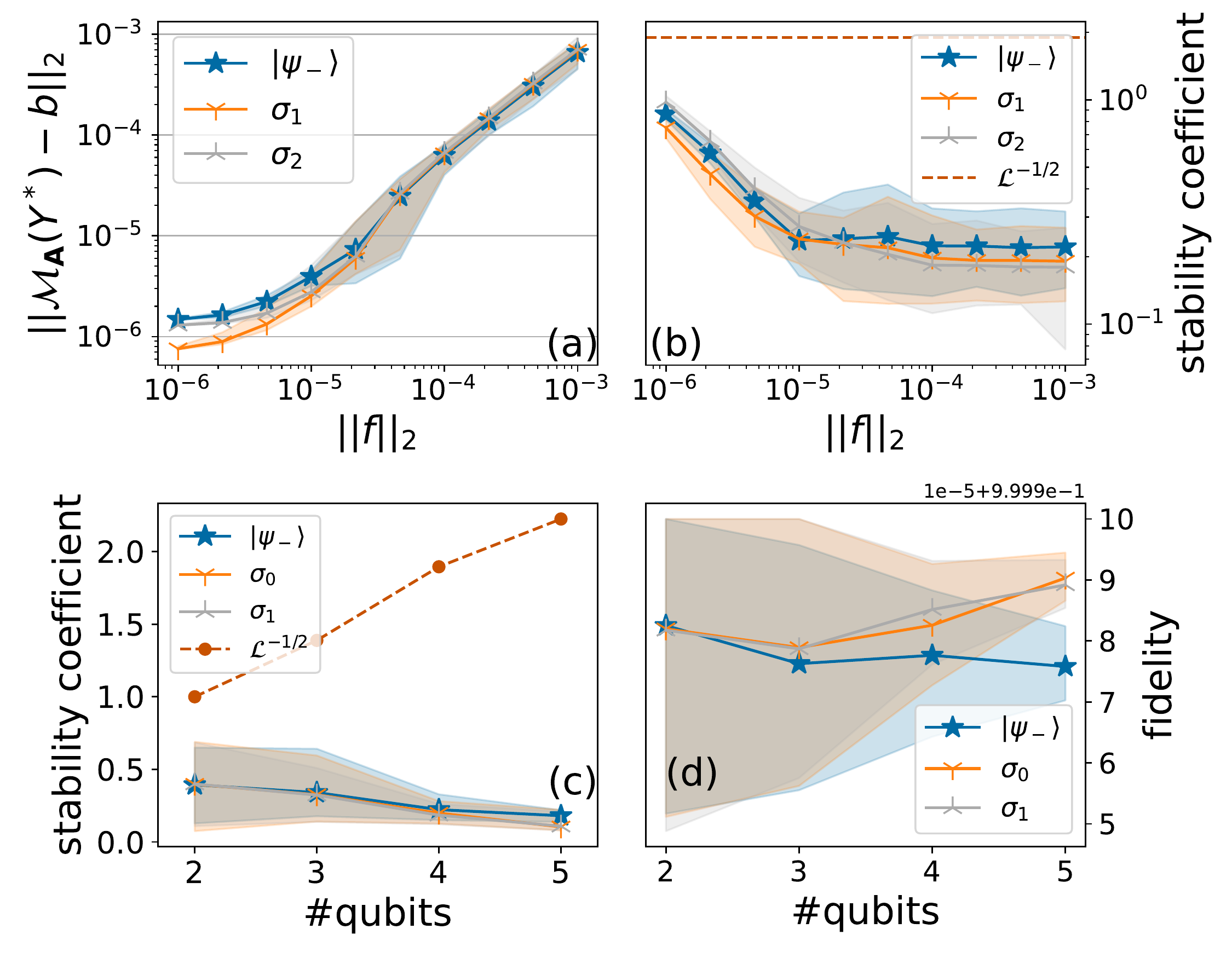}
    \caption{\label{fig:pauli-worst-case}Stability of state recovery using Pauli measurements for $4$ qubits. $|\psi_-\rangle$: the pure state with the negative eigenvalue found in minimizing the loss function (\ref{eq:loss-function}), and $\sigma_1,\sigma_2$: two random pure states. (a)  the recovered error $\lVert\mathcal{M}_A(Y^*)-b\rVert_2$ versus noise rate $\lVert f\rVert_2$, with $n=4$ (b) stability coefficient $\alpha$ versus noise rate $\lVert f\rVert_2$, with $n=4$. (c) stability coefficient $\alpha$ versus number of qubits $n$. (d) fidelity versus the number of qubits $n$. Shaded region denotes (minimum, maximum) range over randomly sampled noise for all subfigures.}
\end{figure}

\appendix
\section{The stability of Pauli measurement for pure-state tomography}
\label{pauli}

Similar calculations are performed on $n$-qubit optimal Pauli measurement schemes found by the search algorithm (see TABLE \ref{tab:pauli-ud}), as shown in FIG. \ref{fig:pauli-worst-case}. Three pure states are considered: $|\psi_-\rangle$, the pure state with the negative eigenvalue found in minimizing the loss function (\ref{eq:loss-function}), and $\sigma_1,\sigma_2$, two random pure states. Different from the 5PB case in Section \ref{polynomial}, the recovered error $\lVert\mathcal{M}_A(Y^*)-b\rVert_2$ and the stability coefficient for $n=4$ qubits, plotted in the subfigure (a) and (b), shows no difference among these three pure states. For a fixed noise rate $\lVert f\rVert_2=10^{-4}$ and various numbers of qubits, no significant difference can be observed among them. More importantly, for such a parameter setting $\lVert f\rVert_2=10^{-4}$, the recovered fidelities for all three pure states are almost $1$, which implies perfect state recovery.

\section{Data Availability}

Both the code and data for our project have been made publicly available. Our open-source repository can be accessed at \cite{github-repo}.

\bibliography{reference.bib}

\begin{thebibliography}{48}%
\makeatletter
\providecommand \@ifxundefined [1]{%
 \@ifx{#1\undefined}
}%
\providecommand \@ifnum [1]{%
 \ifnum #1\expandafter \@firstoftwo
 \else \expandafter \@secondoftwo
 \fi
}%
\providecommand \@ifx [1]{%
 \ifx #1\expandafter \@firstoftwo
 \else \expandafter \@secondoftwo
 \fi
}%
\providecommand \natexlab [1]{#1}%
\providecommand \enquote  [1]{``#1''}%
\providecommand \bibnamefont  [1]{#1}%
\providecommand \bibfnamefont [1]{#1}%
\providecommand \citenamefont [1]{#1}%
\providecommand \href@noop [0]{\@secondoftwo}%
\providecommand \href [0]{\begingroup \@sanitize@url \@href}%
\providecommand \@href[1]{\@@startlink{#1}\@@href}%
\providecommand \@@href[1]{\endgroup#1\@@endlink}%
\providecommand \@sanitize@url [0]{\catcode `\\12\catcode `\$12\catcode
  `\&12\catcode `\#12\catcode `\^12\catcode `\_12\catcode `\%12\relax}%
\providecommand \@@startlink[1]{}%
\providecommand \@@endlink[0]{}%
\providecommand \url  [0]{\begingroup\@sanitize@url \@url }%
\providecommand \@url [1]{\endgroup\@href {#1}{\urlprefix }}%
\providecommand \urlprefix  [0]{URL }%
\providecommand \Eprint [0]{\href }%
\providecommand \doibase [0]{https://doi.org/}%
\providecommand \selectlanguage [0]{\@gobble}%
\providecommand \bibinfo  [0]{\@secondoftwo}%
\providecommand \bibfield  [0]{\@secondoftwo}%
\providecommand \translation [1]{[#1]}%
\providecommand \BibitemOpen [0]{}%
\providecommand \bibitemStop [0]{}%
\providecommand \bibitemNoStop [0]{.\EOS\space}%
\providecommand \EOS [0]{\spacefactor3000\relax}%
\providecommand \BibitemShut  [1]{\csname bibitem#1\endcsname}%
\let\auto@bib@innerbib\@empty
\bibitem [{\citenamefont {Christandl}\ and\ \citenamefont
  {Renner}(2012)}]{christandl2012reliable}%
  \BibitemOpen
  \bibfield  {author} {\bibinfo {author} {\bibfnamefont {M.}~\bibnamefont
  {Christandl}}\ and\ \bibinfo {author} {\bibfnamefont {R.}~\bibnamefont
  {Renner}},\ }\bibfield  {title} {\bibinfo {title} {Reliable quantum state
  tomography},\ }\href {https://doi.org/10.1103/PhysRevLett.109.120403}
  {\bibfield  {journal} {\bibinfo  {journal} {Phys. Rev. Lett.}\ }\textbf
  {\bibinfo {volume} {109}},\ \bibinfo {pages} {120403} (\bibinfo {year}
  {2012})}\BibitemShut {NoStop}%
\bibitem [{\citenamefont {Lvovsky}\ and\ \citenamefont
  {Raymer}(2009)}]{lvovsky2009continuous}%
  \BibitemOpen
  \bibfield  {author} {\bibinfo {author} {\bibfnamefont {A.~I.}\ \bibnamefont
  {Lvovsky}}\ and\ \bibinfo {author} {\bibfnamefont {M.~G.}\ \bibnamefont
  {Raymer}},\ }\bibfield  {title} {\bibinfo {title} {Continuous-variable
  optical quantum-state tomography},\ }\href
  {https://doi.org/10.1103/RevModPhys.81.299} {\bibfield  {journal} {\bibinfo
  {journal} {Rev. Mod. Phys.}\ }\textbf {\bibinfo {volume} {81}},\ \bibinfo
  {pages} {299} (\bibinfo {year} {2009})}\BibitemShut {NoStop}%
\bibitem [{\citenamefont {Thew}\ \emph {et~al.}(2002)\citenamefont {Thew},
  \citenamefont {Nemoto}, \citenamefont {White},\ and\ \citenamefont
  {Munro}}]{thew2002qudit}%
  \BibitemOpen
  \bibfield  {author} {\bibinfo {author} {\bibfnamefont {R.~T.}\ \bibnamefont
  {Thew}}, \bibinfo {author} {\bibfnamefont {K.}~\bibnamefont {Nemoto}},
  \bibinfo {author} {\bibfnamefont {A.~G.}\ \bibnamefont {White}},\ and\
  \bibinfo {author} {\bibfnamefont {W.~J.}\ \bibnamefont {Munro}},\ }\bibfield
  {title} {\bibinfo {title} {Qudit quantum-state tomography},\ }\href
  {https://doi.org/10.1103/PhysRevA.66.012303} {\bibfield  {journal} {\bibinfo
  {journal} {Phys. Rev. A}\ }\textbf {\bibinfo {volume} {66}},\ \bibinfo
  {pages} {012303} (\bibinfo {year} {2002})}\BibitemShut {NoStop}%
\bibitem [{\citenamefont {Qi}\ \emph {et~al.}(2013)\citenamefont {Qi},
  \citenamefont {Hou}, \citenamefont {Li}, \citenamefont {Dong}, \citenamefont
  {Xiang},\ and\ \citenamefont {Guo}}]{qi2013quantum}%
  \BibitemOpen
  \bibfield  {author} {\bibinfo {author} {\bibfnamefont {B.}~\bibnamefont
  {Qi}}, \bibinfo {author} {\bibfnamefont {Z.}~\bibnamefont {Hou}}, \bibinfo
  {author} {\bibfnamefont {L.}~\bibnamefont {Li}}, \bibinfo {author}
  {\bibfnamefont {D.}~\bibnamefont {Dong}}, \bibinfo {author} {\bibfnamefont
  {G.}~\bibnamefont {Xiang}},\ and\ \bibinfo {author} {\bibfnamefont
  {G.}~\bibnamefont {Guo}},\ }\bibfield  {title} {\bibinfo {title} {Quantum
  {State} {Tomography} via {Linear} {Regression} {Estimation}},\ }\href
  {https://doi.org/10.1038/srep03496} {\bibfield  {journal} {\bibinfo
  {journal} {Sci. Rep.}\ }\textbf {\bibinfo {volume} {3}},\ \bibinfo {pages}
  {3496} (\bibinfo {year} {2013})}\BibitemShut {NoStop}%
\bibitem [{par(2004)}]{paris2004quantum}%
  \BibitemOpen
  \href {https://doi.org/10.1007/b98673} {\emph {\bibinfo {title} {Quantum
  State Estimation}}}\ (\bibinfo  {publisher} {Springer Berlin Heidelberg},\
  \bibinfo {year} {2004})\BibitemShut {NoStop}%
\bibitem [{\citenamefont {Steane}(1998)}]{steane1998quantum}%
  \BibitemOpen
  \bibfield  {author} {\bibinfo {author} {\bibfnamefont {A.}~\bibnamefont
  {Steane}},\ }\bibfield  {title} {\bibinfo {title} {Quantum computing},\
  }\href {https://doi.org/10.1088/0034-4885/61/2/002} {\bibfield  {journal}
  {\bibinfo  {journal} {Rep. Prog. Phys}\ }\textbf {\bibinfo {volume} {61}},\
  \bibinfo {pages} {117} (\bibinfo {year} {1998})}\BibitemShut {NoStop}%
\bibitem [{\citenamefont {O'Brien}(2007)}]{o2007optical}%
  \BibitemOpen
  \bibfield  {author} {\bibinfo {author} {\bibfnamefont {J.~L.}\ \bibnamefont
  {O'Brien}},\ }\bibfield  {title} {\bibinfo {title} {Optical quantum
  computing},\ }\href {https://doi.org/10.1126/science.1142892} {\bibfield
  {journal} {\bibinfo  {journal} {Science}\ }\textbf {\bibinfo {volume}
  {318}},\ \bibinfo {pages} {1567} (\bibinfo {year} {2007})},\ \Eprint
  {https://arxiv.org/abs/https://www.science.org/doi/pdf/10.1126/science.1142892}
  {https://www.science.org/doi/pdf/10.1126/science.1142892} \BibitemShut
  {NoStop}%
\bibitem [{\citenamefont {Preskill}(2018)}]{preskill2018quantum}%
  \BibitemOpen
  \bibfield  {author} {\bibinfo {author} {\bibfnamefont {J.}~\bibnamefont
  {Preskill}},\ }\bibfield  {title} {\bibinfo {title} {Quantum {C}omputing in
  the {NISQ} era and beyond},\ }\href
  {https://doi.org/10.22331/q-2018-08-06-79} {\bibfield  {journal} {\bibinfo
  {journal} {{Quantum}}\ }\textbf {\bibinfo {volume} {2}},\ \bibinfo {pages}
  {79} (\bibinfo {year} {2018})}\BibitemShut {NoStop}%
\bibitem [{\citenamefont {Gisin}\ and\ \citenamefont
  {Thew}(2007)}]{gisin2007quantum}%
  \BibitemOpen
  \bibfield  {author} {\bibinfo {author} {\bibfnamefont {N.}~\bibnamefont
  {Gisin}}\ and\ \bibinfo {author} {\bibfnamefont {R.}~\bibnamefont {Thew}},\
  }\bibfield  {title} {\bibinfo {title} {Quantum communication},\ }\href
  {https://doi.org/10.1038/nphoton.2007.22} {\bibfield  {journal} {\bibinfo
  {journal} {Nat. Photonics}\ }\textbf {\bibinfo {volume} {1}},\ \bibinfo
  {pages} {165} (\bibinfo {year} {2007})}\BibitemShut {NoStop}%
\bibitem [{\citenamefont {Cozzolino}\ \emph {et~al.}(2019)\citenamefont
  {Cozzolino}, \citenamefont {Da~Lio}, \citenamefont {Bacco},\ and\
  \citenamefont {Oxenløwe}}]{cozzolino2019high}%
  \BibitemOpen
  \bibfield  {author} {\bibinfo {author} {\bibfnamefont {D.}~\bibnamefont
  {Cozzolino}}, \bibinfo {author} {\bibfnamefont {B.}~\bibnamefont {Da~Lio}},
  \bibinfo {author} {\bibfnamefont {D.}~\bibnamefont {Bacco}},\ and\ \bibinfo
  {author} {\bibfnamefont {L.~K.}\ \bibnamefont {Oxenløwe}},\ }\bibfield
  {title} {\bibinfo {title} {High-dimensional quantum communication: Benefits,
  progress, and future challenges},\ }\href
  {https://doi.org/https://doi.org/10.1002/qute.201900038} {\bibfield
  {journal} {\bibinfo  {journal} {Adv. Quantum Technol.}\ }\textbf {\bibinfo
  {volume} {2}},\ \bibinfo {pages} {1900038} (\bibinfo {year} {2019})},\
  \Eprint
  {https://arxiv.org/abs/https://onlinelibrary.wiley.com/doi/pdf/10.1002/qute.201900038}
  {https://onlinelibrary.wiley.com/doi/pdf/10.1002/qute.201900038} \BibitemShut
  {NoStop}%
\bibitem [{\citenamefont {Brassard}(2003)}]{brassard2003quantum}%
  \BibitemOpen
  \bibfield  {author} {\bibinfo {author} {\bibfnamefont {G.}~\bibnamefont
  {Brassard}},\ }\bibfield  {title} {\bibinfo {title} {Quantum {Communication}
  {Complexity}},\ }\href {https://doi.org/10.1023/A:1026009100467} {\bibfield
  {journal} {\bibinfo  {journal} {Found. Phys.}\ }\textbf {\bibinfo {volume}
  {33}},\ \bibinfo {pages} {1593} (\bibinfo {year} {2003})}\BibitemShut
  {NoStop}%
\bibitem [{\citenamefont {Gisin}\ \emph {et~al.}(2002)\citenamefont {Gisin},
  \citenamefont {Ribordy}, \citenamefont {Tittel},\ and\ \citenamefont
  {Zbinden}}]{gisin2002quantum}%
  \BibitemOpen
  \bibfield  {author} {\bibinfo {author} {\bibfnamefont {N.}~\bibnamefont
  {Gisin}}, \bibinfo {author} {\bibfnamefont {G.}~\bibnamefont {Ribordy}},
  \bibinfo {author} {\bibfnamefont {W.}~\bibnamefont {Tittel}},\ and\ \bibinfo
  {author} {\bibfnamefont {H.}~\bibnamefont {Zbinden}},\ }\bibfield  {title}
  {\bibinfo {title} {Quantum cryptography},\ }\href
  {https://doi.org/10.1103/RevModPhys.74.145} {\bibfield  {journal} {\bibinfo
  {journal} {Rev. Mod. Phys.}\ }\textbf {\bibinfo {volume} {74}},\ \bibinfo
  {pages} {145} (\bibinfo {year} {2002})}\BibitemShut {NoStop}%
\bibitem [{\citenamefont {Pirandola}\ \emph {et~al.}(2020)\citenamefont
  {Pirandola}, \citenamefont {Andersen}, \citenamefont {Banchi}, \citenamefont
  {Berta}, \citenamefont {Bunandar}, \citenamefont {Colbeck}, \citenamefont
  {Englund}, \citenamefont {Gehring}, \citenamefont {Lupo}, \citenamefont
  {Ottaviani}, \citenamefont {Pereira}, \citenamefont {Razavi}, \citenamefont
  {Shaari}, \citenamefont {Tomamichel}, \citenamefont {Usenko}, \citenamefont
  {Vallone}, \citenamefont {Villoresi},\ and\ \citenamefont
  {Wallden}}]{pirandola2020advances}%
  \BibitemOpen
  \bibfield  {author} {\bibinfo {author} {\bibfnamefont {S.}~\bibnamefont
  {Pirandola}}, \bibinfo {author} {\bibfnamefont {U.~L.}\ \bibnamefont
  {Andersen}}, \bibinfo {author} {\bibfnamefont {L.}~\bibnamefont {Banchi}},
  \bibinfo {author} {\bibfnamefont {M.}~\bibnamefont {Berta}}, \bibinfo
  {author} {\bibfnamefont {D.}~\bibnamefont {Bunandar}}, \bibinfo {author}
  {\bibfnamefont {R.}~\bibnamefont {Colbeck}}, \bibinfo {author} {\bibfnamefont
  {D.}~\bibnamefont {Englund}}, \bibinfo {author} {\bibfnamefont
  {T.}~\bibnamefont {Gehring}}, \bibinfo {author} {\bibfnamefont
  {C.}~\bibnamefont {Lupo}}, \bibinfo {author} {\bibfnamefont {C.}~\bibnamefont
  {Ottaviani}}, \bibinfo {author} {\bibfnamefont {J.~L.}\ \bibnamefont
  {Pereira}}, \bibinfo {author} {\bibfnamefont {M.}~\bibnamefont {Razavi}},
  \bibinfo {author} {\bibfnamefont {J.~S.}\ \bibnamefont {Shaari}}, \bibinfo
  {author} {\bibfnamefont {M.}~\bibnamefont {Tomamichel}}, \bibinfo {author}
  {\bibfnamefont {V.~C.}\ \bibnamefont {Usenko}}, \bibinfo {author}
  {\bibfnamefont {G.}~\bibnamefont {Vallone}}, \bibinfo {author} {\bibfnamefont
  {P.}~\bibnamefont {Villoresi}},\ and\ \bibinfo {author} {\bibfnamefont
  {P.}~\bibnamefont {Wallden}},\ }\bibfield  {title} {\bibinfo {title}
  {Advances in quantum cryptography},\ }\href
  {https://doi.org/10.1364/AOP.361502} {\bibfield  {journal} {\bibinfo
  {journal} {Adv. Opt. Photon.}\ }\textbf {\bibinfo {volume} {12}},\ \bibinfo
  {pages} {1012} (\bibinfo {year} {2020})}\BibitemShut {NoStop}%
\bibitem [{\citenamefont {Bennett}\ \emph {et~al.}(1992)\citenamefont
  {Bennett}, \citenamefont {Brassard},\ and\ \citenamefont
  {Ekert}}]{bennett1992quantum}%
  \BibitemOpen
  \bibfield  {author} {\bibinfo {author} {\bibfnamefont {C.~H.}\ \bibnamefont
  {Bennett}}, \bibinfo {author} {\bibfnamefont {G.}~\bibnamefont {Brassard}},\
  and\ \bibinfo {author} {\bibfnamefont {A.~K.}\ \bibnamefont {Ekert}},\
  }\bibfield  {title} {\bibinfo {title} {Quantum cryptography},\ }\href
  {http://www.jstor.org/stable/24939253} {\bibfield  {journal} {\bibinfo
  {journal} {Sci. Am.}\ }\textbf {\bibinfo {volume} {267}},\ \bibinfo {pages}
  {50} (\bibinfo {year} {1992})}\BibitemShut {NoStop}%
\bibitem [{\citenamefont {Renes}\ \emph {et~al.}(2004)\citenamefont {Renes},
  \citenamefont {Blume-Kohout}, \citenamefont {Scott},\ and\ \citenamefont
  {Caves}}]{renes2004symmetric}%
  \BibitemOpen
  \bibfield  {author} {\bibinfo {author} {\bibfnamefont {J.~M.}\ \bibnamefont
  {Renes}}, \bibinfo {author} {\bibfnamefont {R.}~\bibnamefont {Blume-Kohout}},
  \bibinfo {author} {\bibfnamefont {A.~J.}\ \bibnamefont {Scott}},\ and\
  \bibinfo {author} {\bibfnamefont {C.~M.}\ \bibnamefont {Caves}},\ }\bibfield
  {title} {\bibinfo {title} {{Symmetric informationally complete quantum
  measurements}},\ }\href {https://doi.org/10.1063/1.1737053} {\bibfield
  {journal} {\bibinfo  {journal} {J. Math. Phys.}\ }\textbf {\bibinfo {volume}
  {45}},\ \bibinfo {pages} {2171} (\bibinfo {year} {2004})},\ \Eprint
  {https://arxiv.org/abs/https://pubs.aip.org/aip/jmp/article-pdf/45/6/2171/8173125/2171\_1\_online.pdf}
  {https://pubs.aip.org/aip/jmp/article-pdf/45/6/2171/8173125/2171\_1\_online.pdf}
  \BibitemShut {NoStop}%
\bibitem [{\citenamefont {Wootters}\ and\ \citenamefont
  {Fields}(1989)}]{wootters1989optimal}%
  \BibitemOpen
  \bibfield  {author} {\bibinfo {author} {\bibfnamefont {W.~K.}\ \bibnamefont
  {Wootters}}\ and\ \bibinfo {author} {\bibfnamefont {B.~D.}\ \bibnamefont
  {Fields}},\ }\bibfield  {title} {\bibinfo {title} {Optimal
  state-determination by mutually unbiased measurements},\ }\href
  {https://doi.org/https://doi.org/10.1016/0003-4916(89)90322-9} {\bibfield
  {journal} {\bibinfo  {journal} {Ann. Phys.}\ }\textbf {\bibinfo {volume}
  {191}},\ \bibinfo {pages} {363} (\bibinfo {year} {1989})}\BibitemShut
  {NoStop}%
\bibitem [{\citenamefont {de~Burgh}\ \emph {et~al.}(2008)\citenamefont
  {de~Burgh}, \citenamefont {Langford}, \citenamefont {Doherty},\ and\
  \citenamefont {Gilchrist}}]{de2008choice}%
  \BibitemOpen
  \bibfield  {author} {\bibinfo {author} {\bibfnamefont {M.~D.}\ \bibnamefont
  {de~Burgh}}, \bibinfo {author} {\bibfnamefont {N.~K.}\ \bibnamefont
  {Langford}}, \bibinfo {author} {\bibfnamefont {A.~C.}\ \bibnamefont
  {Doherty}},\ and\ \bibinfo {author} {\bibfnamefont {A.}~\bibnamefont
  {Gilchrist}},\ }\bibfield  {title} {\bibinfo {title} {Choice of measurement
  sets in qubit tomography},\ }\href
  {https://doi.org/10.1103/PhysRevA.78.052122} {\bibfield  {journal} {\bibinfo
  {journal} {Phys. Rev. A}\ }\textbf {\bibinfo {volume} {78}},\ \bibinfo
  {pages} {052122} (\bibinfo {year} {2008})}\BibitemShut {NoStop}%
\bibitem [{\citenamefont {Adamson}\ and\ \citenamefont
  {Steinberg}(2010)}]{adamson2010improving}%
  \BibitemOpen
  \bibfield  {author} {\bibinfo {author} {\bibfnamefont {R.~B.~A.}\
  \bibnamefont {Adamson}}\ and\ \bibinfo {author} {\bibfnamefont {A.~M.}\
  \bibnamefont {Steinberg}},\ }\bibfield  {title} {\bibinfo {title} {Improving
  quantum state estimation with mutually unbiased bases},\ }\href
  {https://doi.org/10.1103/PhysRevLett.105.030406} {\bibfield  {journal}
  {\bibinfo  {journal} {Phys. Rev. Lett.}\ }\textbf {\bibinfo {volume} {105}},\
  \bibinfo {pages} {030406} (\bibinfo {year} {2010})}\BibitemShut {NoStop}%
\bibitem [{\citenamefont {Pauli}(1933)}]{pauli1933allgemeinen}%
  \BibitemOpen
  \bibfield  {author} {\bibinfo {author} {\bibfnamefont {W.}~\bibnamefont
  {Pauli}},\ }\bibfield  {title} {\bibinfo {title} {Die allgemeinen prinzipien
  der wellenmechanik},\ }in\ \href
  {https://doi.org/10.1007/978-3-642-52619-0_2} {\emph {\bibinfo {booktitle}
  {Quantentheorie}}}\ (\bibinfo  {publisher} {Springer Berlin Heidelberg},\
  \bibinfo {year} {1933})\ pp.\ \bibinfo {pages} {83--272}\BibitemShut
  {NoStop}%
\bibitem [{\citenamefont {Xin}\ \emph {et~al.}(2020)\citenamefont {Xin},
  \citenamefont {Nie}, \citenamefont {Kong}, \citenamefont {Wen}, \citenamefont
  {Lu},\ and\ \citenamefont {Li}}]{xin2020quantum}%
  \BibitemOpen
  \bibfield  {author} {\bibinfo {author} {\bibfnamefont {T.}~\bibnamefont
  {Xin}}, \bibinfo {author} {\bibfnamefont {X.}~\bibnamefont {Nie}}, \bibinfo
  {author} {\bibfnamefont {X.}~\bibnamefont {Kong}}, \bibinfo {author}
  {\bibfnamefont {J.}~\bibnamefont {Wen}}, \bibinfo {author} {\bibfnamefont
  {D.}~\bibnamefont {Lu}},\ and\ \bibinfo {author} {\bibfnamefont
  {J.}~\bibnamefont {Li}},\ }\bibfield  {title} {\bibinfo {title} {Quantum pure
  state tomography via variational hybrid quantum-classical method},\ }\href
  {https://doi.org/10.1103/PhysRevApplied.13.024013} {\bibfield  {journal}
  {\bibinfo  {journal} {Phys. Rev. Appl.}\ }\textbf {\bibinfo {volume} {13}},\
  \bibinfo {pages} {024013} (\bibinfo {year} {2020})}\BibitemShut {NoStop}%
\bibitem [{\citenamefont {Ma}\ \emph {et~al.}(2016)\citenamefont {Ma},
  \citenamefont {Jackson}, \citenamefont {Zhou}, \citenamefont {Chen},
  \citenamefont {Lu}, \citenamefont {Mazurek}, \citenamefont {Fisher},
  \citenamefont {Peng}, \citenamefont {Kribs}, \citenamefont {Resch},
  \citenamefont {Ji}, \citenamefont {Zeng},\ and\ \citenamefont
  {Laflamme}}]{ma2016pure}%
  \BibitemOpen
  \bibfield  {author} {\bibinfo {author} {\bibfnamefont {X.}~\bibnamefont
  {Ma}}, \bibinfo {author} {\bibfnamefont {T.}~\bibnamefont {Jackson}},
  \bibinfo {author} {\bibfnamefont {H.}~\bibnamefont {Zhou}}, \bibinfo {author}
  {\bibfnamefont {J.}~\bibnamefont {Chen}}, \bibinfo {author} {\bibfnamefont
  {D.}~\bibnamefont {Lu}}, \bibinfo {author} {\bibfnamefont {M.~D.}\
  \bibnamefont {Mazurek}}, \bibinfo {author} {\bibfnamefont {K.~A.~G.}\
  \bibnamefont {Fisher}}, \bibinfo {author} {\bibfnamefont {X.}~\bibnamefont
  {Peng}}, \bibinfo {author} {\bibfnamefont {D.}~\bibnamefont {Kribs}},
  \bibinfo {author} {\bibfnamefont {K.~J.}\ \bibnamefont {Resch}}, \bibinfo
  {author} {\bibfnamefont {Z.}~\bibnamefont {Ji}}, \bibinfo {author}
  {\bibfnamefont {B.}~\bibnamefont {Zeng}},\ and\ \bibinfo {author}
  {\bibfnamefont {R.}~\bibnamefont {Laflamme}},\ }\bibfield  {title} {\bibinfo
  {title} {Pure-state tomography with the expectation value of pauli
  operators},\ }\href {https://doi.org/10.1103/PhysRevA.93.032140} {\bibfield
  {journal} {\bibinfo  {journal} {Phys. Rev. A}\ }\textbf {\bibinfo {volume}
  {93}},\ \bibinfo {pages} {032140} (\bibinfo {year} {2016})}\BibitemShut
  {NoStop}%
\bibitem [{\citenamefont {Sosa-Martinez}\ \emph {et~al.}(2017)\citenamefont
  {Sosa-Martinez}, \citenamefont {Lysne}, \citenamefont {Baldwin},
  \citenamefont {Kalev}, \citenamefont {Deutsch},\ and\ \citenamefont
  {Jessen}}]{sosa2017experimental}%
  \BibitemOpen
  \bibfield  {author} {\bibinfo {author} {\bibfnamefont {H.}~\bibnamefont
  {Sosa-Martinez}}, \bibinfo {author} {\bibfnamefont {N.~K.}\ \bibnamefont
  {Lysne}}, \bibinfo {author} {\bibfnamefont {C.~H.}\ \bibnamefont {Baldwin}},
  \bibinfo {author} {\bibfnamefont {A.}~\bibnamefont {Kalev}}, \bibinfo
  {author} {\bibfnamefont {I.~H.}\ \bibnamefont {Deutsch}},\ and\ \bibinfo
  {author} {\bibfnamefont {P.~S.}\ \bibnamefont {Jessen}},\ }\bibfield  {title}
  {\bibinfo {title} {Experimental study of optimal measurements for quantum
  state tomography},\ }\href {https://doi.org/10.1103/PhysRevLett.119.150401}
  {\bibfield  {journal} {\bibinfo  {journal} {Phys. Rev. Lett.}\ }\textbf
  {\bibinfo {volume} {119}},\ \bibinfo {pages} {150401} (\bibinfo {year}
  {2017})}\BibitemShut {NoStop}%
\bibitem [{\citenamefont {Liu}\ \emph {et~al.}(2012)\citenamefont {Liu},
  \citenamefont {Zhang}, \citenamefont {Liu}, \citenamefont {Chen},\ and\
  \citenamefont {Yuan}}]{liu2012experimental}%
  \BibitemOpen
  \bibfield  {author} {\bibinfo {author} {\bibfnamefont {W.-T.}\ \bibnamefont
  {Liu}}, \bibinfo {author} {\bibfnamefont {T.}~\bibnamefont {Zhang}}, \bibinfo
  {author} {\bibfnamefont {J.-Y.}\ \bibnamefont {Liu}}, \bibinfo {author}
  {\bibfnamefont {P.-X.}\ \bibnamefont {Chen}},\ and\ \bibinfo {author}
  {\bibfnamefont {J.-M.}\ \bibnamefont {Yuan}},\ }\bibfield  {title} {\bibinfo
  {title} {Experimental quantum state tomography via compressed sampling},\
  }\href {https://doi.org/10.1103/PhysRevLett.108.170403} {\bibfield  {journal}
  {\bibinfo  {journal} {Phys. Rev. Lett.}\ }\textbf {\bibinfo {volume} {108}},\
  \bibinfo {pages} {170403} (\bibinfo {year} {2012})}\BibitemShut {NoStop}%
\bibitem [{\citenamefont {Heinosaari}\ \emph {et~al.}(2013)\citenamefont
  {Heinosaari}, \citenamefont {Mazzarella},\ and\ \citenamefont
  {Wolf}}]{heinosaari2013quantum}%
  \BibitemOpen
  \bibfield  {author} {\bibinfo {author} {\bibfnamefont {T.}~\bibnamefont
  {Heinosaari}}, \bibinfo {author} {\bibfnamefont {L.}~\bibnamefont
  {Mazzarella}},\ and\ \bibinfo {author} {\bibfnamefont {M.~M.}\ \bibnamefont
  {Wolf}},\ }\bibfield  {title} {\bibinfo {title} {Quantum {Tomography} under
  {Prior} {Information}},\ }\href {https://doi.org/10.1007/s00220-013-1671-8}
  {\bibfield  {journal} {\bibinfo  {journal} {Commun. Math. Phys.}\ }\textbf
  {\bibinfo {volume} {318}},\ \bibinfo {pages} {355} (\bibinfo {year}
  {2013})}\BibitemShut {NoStop}%
\bibitem [{\citenamefont {Carmeli}\ \emph {et~al.}(2014)\citenamefont
  {Carmeli}, \citenamefont {Heinosaari}, \citenamefont {Schultz},\ and\
  \citenamefont {Toigo}}]{carmeli2014tasks}%
  \BibitemOpen
  \bibfield  {author} {\bibinfo {author} {\bibfnamefont {C.}~\bibnamefont
  {Carmeli}}, \bibinfo {author} {\bibfnamefont {T.}~\bibnamefont {Heinosaari}},
  \bibinfo {author} {\bibfnamefont {J.}~\bibnamefont {Schultz}},\ and\ \bibinfo
  {author} {\bibfnamefont {A.}~\bibnamefont {Toigo}},\ }\bibfield  {title}
  {\bibinfo {title} {Tasks and premises in quantum state determination},\
  }\href {https://doi.org/10.1088/1751-8113/47/7/075302} {\bibfield  {journal}
  {\bibinfo  {journal} {J. Phys. A: Math. Theor.}\ }\textbf {\bibinfo {volume}
  {47}},\ \bibinfo {pages} {075302} (\bibinfo {year} {2014})}\BibitemShut
  {NoStop}%
\bibitem [{\citenamefont {Kech}\ and\ \citenamefont
  {Wolf}(2016)}]{kech2017constrained}%
  \BibitemOpen
  \bibfield  {author} {\bibinfo {author} {\bibfnamefont {M.}~\bibnamefont
  {Kech}}\ and\ \bibinfo {author} {\bibfnamefont {M.~M.}\ \bibnamefont
  {Wolf}},\ }\bibfield  {title} {\bibinfo {title} {Constrained quantum
  tomography of semi-algebraic sets with applications to low-rank matrix
  recovery},\ }\href {https://doi.org/10.1093/imaiai/iaw019} {\bibfield
  {journal} {\bibinfo  {journal} {Inf. Inference J. IMA}\ ,\ \bibinfo {pages}
  {iaw019}} (\bibinfo {year} {2016})}\BibitemShut {NoStop}%
\bibitem [{\citenamefont {Baldwin}\ \emph {et~al.}(2016)\citenamefont
  {Baldwin}, \citenamefont {Deutsch},\ and\ \citenamefont
  {Kalev}}]{baldwin2016strictly}%
  \BibitemOpen
  \bibfield  {author} {\bibinfo {author} {\bibfnamefont {C.~H.}\ \bibnamefont
  {Baldwin}}, \bibinfo {author} {\bibfnamefont {I.~H.}\ \bibnamefont
  {Deutsch}},\ and\ \bibinfo {author} {\bibfnamefont {A.}~\bibnamefont
  {Kalev}},\ }\bibfield  {title} {\bibinfo {title} {Strictly-complete
  measurements for bounded-rank quantum-state tomography},\ }\href
  {https://doi.org/10.1103/PhysRevA.93.052105} {\bibfield  {journal} {\bibinfo
  {journal} {Phys. Rev. A}\ }\textbf {\bibinfo {volume} {93}},\ \bibinfo
  {pages} {052105} (\bibinfo {year} {2016})}\BibitemShut {NoStop}%
\bibitem [{\citenamefont {Cramer}\ \emph {et~al.}(2010)\citenamefont {Cramer},
  \citenamefont {Plenio}, \citenamefont {Flammia}, \citenamefont {Somma},
  \citenamefont {Gross}, \citenamefont {Bartlett}, \citenamefont
  {Landon-Cardinal}, \citenamefont {Poulin},\ and\ \citenamefont
  {Liu}}]{cramer2010efficient}%
  \BibitemOpen
  \bibfield  {author} {\bibinfo {author} {\bibfnamefont {M.}~\bibnamefont
  {Cramer}}, \bibinfo {author} {\bibfnamefont {M.~B.}\ \bibnamefont {Plenio}},
  \bibinfo {author} {\bibfnamefont {S.~T.}\ \bibnamefont {Flammia}}, \bibinfo
  {author} {\bibfnamefont {R.}~\bibnamefont {Somma}}, \bibinfo {author}
  {\bibfnamefont {D.}~\bibnamefont {Gross}}, \bibinfo {author} {\bibfnamefont
  {S.~D.}\ \bibnamefont {Bartlett}}, \bibinfo {author} {\bibfnamefont
  {O.}~\bibnamefont {Landon-Cardinal}}, \bibinfo {author} {\bibfnamefont
  {D.}~\bibnamefont {Poulin}},\ and\ \bibinfo {author} {\bibfnamefont {Y.-K.}\
  \bibnamefont {Liu}},\ }\bibfield  {title} {\bibinfo {title} {Efficient
  quantum state tomography},\ }\href {https://doi.org/10.1038/ncomms1147}
  {\bibfield  {journal} {\bibinfo  {journal} {Nat. Commun.}\ }\textbf {\bibinfo
  {volume} {1}},\ \bibinfo {pages} {149} (\bibinfo {year} {2010})}\BibitemShut
  {NoStop}%
\bibitem [{\citenamefont {Klyachko}(2006)}]{Klyachko_2006}%
  \BibitemOpen
  \bibfield  {author} {\bibinfo {author} {\bibfnamefont {A.~A.}\ \bibnamefont
  {Klyachko}},\ }\bibfield  {title} {\bibinfo {title} {Quantum marginal problem
  and n-representability},\ }\href {https://doi.org/10.1088/1742-6596/36/1/014}
  {\bibfield  {journal} {\bibinfo  {journal} {Journal of Physics: Conference
  Series}\ }\textbf {\bibinfo {volume} {36}},\ \bibinfo {pages} {72–86}
  (\bibinfo {year} {2006})}\BibitemShut {NoStop}%
\bibitem [{\citenamefont {Bae}\ and\ \citenamefont {Kwek}(2015)}]{Bae_2015}%
  \BibitemOpen
  \bibfield  {author} {\bibinfo {author} {\bibfnamefont {J.}~\bibnamefont
  {Bae}}\ and\ \bibinfo {author} {\bibfnamefont {L.-C.}\ \bibnamefont {Kwek}},\
  }\bibfield  {title} {\bibinfo {title} {Quantum state discrimination and its
  applications},\ }\href {https://doi.org/10.1088/1751-8113/48/8/083001}
  {\bibfield  {journal} {\bibinfo  {journal} {Journal of Physics A:
  Mathematical and Theoretical}\ }\textbf {\bibinfo {volume} {48}},\ \bibinfo
  {pages} {083001} (\bibinfo {year} {2015})}\BibitemShut {NoStop}%
\bibitem [{\citenamefont {Flammia}\ \emph {et~al.}(2005)\citenamefont
  {Flammia}, \citenamefont {Silberfarb},\ and\ \citenamefont
  {Caves}}]{flammia2005minimal}%
  \BibitemOpen
  \bibfield  {author} {\bibinfo {author} {\bibfnamefont {S.~T.}\ \bibnamefont
  {Flammia}}, \bibinfo {author} {\bibfnamefont {A.}~\bibnamefont
  {Silberfarb}},\ and\ \bibinfo {author} {\bibfnamefont {C.~M.}\ \bibnamefont
  {Caves}},\ }\bibfield  {title} {\bibinfo {title} {Minimal {Informationally}
  {Complete} {Measurements} for {Pure} {States}},\ }\href
  {https://doi.org/10.1007/s10701-005-8658-z} {\bibfield  {journal} {\bibinfo
  {journal} {Found. Phys.}\ }\textbf {\bibinfo {volume} {35}},\ \bibinfo
  {pages} {1985} (\bibinfo {year} {2005})}\BibitemShut {NoStop}%
\bibitem [{\citenamefont {Scott}(2006)}]{scott2006tight}%
  \BibitemOpen
  \bibfield  {author} {\bibinfo {author} {\bibfnamefont {A.~J.}\ \bibnamefont
  {Scott}},\ }\bibfield  {title} {\bibinfo {title} {Tight informationally
  complete quantum measurements},\ }\href
  {https://doi.org/10.1088/0305-4470/39/43/009} {\bibfield  {journal} {\bibinfo
   {journal} {J. Phys. A: Math. Gen.}\ }\textbf {\bibinfo {volume} {39}},\
  \bibinfo {pages} {13507} (\bibinfo {year} {2006})}\BibitemShut {NoStop}%
\bibitem [{\citenamefont {Chen}\ \emph {et~al.}(2013)\citenamefont {Chen},
  \citenamefont {Dawkins}, \citenamefont {Ji}, \citenamefont {Johnston},
  \citenamefont {Kribs}, \citenamefont {Shultz},\ and\ \citenamefont
  {Zeng}}]{PhysRevA.88.012109}%
  \BibitemOpen
  \bibfield  {author} {\bibinfo {author} {\bibfnamefont {J.}~\bibnamefont
  {Chen}}, \bibinfo {author} {\bibfnamefont {H.}~\bibnamefont {Dawkins}},
  \bibinfo {author} {\bibfnamefont {Z.}~\bibnamefont {Ji}}, \bibinfo {author}
  {\bibfnamefont {N.}~\bibnamefont {Johnston}}, \bibinfo {author}
  {\bibfnamefont {D.}~\bibnamefont {Kribs}}, \bibinfo {author} {\bibfnamefont
  {F.}~\bibnamefont {Shultz}},\ and\ \bibinfo {author} {\bibfnamefont
  {B.}~\bibnamefont {Zeng}},\ }\bibfield  {title} {\bibinfo {title} {Uniqueness
  of quantum states compatible with given measurement results},\ }\href
  {https://doi.org/10.1103/PhysRevA.88.012109} {\bibfield  {journal} {\bibinfo
  {journal} {Phys. Rev. A}\ }\textbf {\bibinfo {volume} {88}},\ \bibinfo
  {pages} {012109} (\bibinfo {year} {2013})}\BibitemShut {NoStop}%
\bibitem [{\citenamefont {Carmeli}\ \emph {et~al.}(2015)\citenamefont
  {Carmeli}, \citenamefont {Heinosaari}, \citenamefont {Schultz},\ and\
  \citenamefont {Toigo}}]{carmeli2015many}%
  \BibitemOpen
  \bibfield  {author} {\bibinfo {author} {\bibfnamefont {C.}~\bibnamefont
  {Carmeli}}, \bibinfo {author} {\bibfnamefont {T.}~\bibnamefont {Heinosaari}},
  \bibinfo {author} {\bibfnamefont {J.}~\bibnamefont {Schultz}},\ and\ \bibinfo
  {author} {\bibfnamefont {A.}~\bibnamefont {Toigo}},\ }\bibfield  {title}
  {\bibinfo {title} {How many orthonormal bases are needed to distinguish all
  pure quantum states?},\ }\href {https://doi.org/10.1140/epjd/e2015-60230-5}
  {\bibfield  {journal} {\bibinfo  {journal} {Eur. Phys. J. D}\ }\textbf
  {\bibinfo {volume} {69}},\ \bibinfo {pages} {179} (\bibinfo {year}
  {2015})}\BibitemShut {NoStop}%
\bibitem [{\citenamefont {Carmeli}\ \emph {et~al.}(2016)\citenamefont
  {Carmeli}, \citenamefont {Heinosaari}, \citenamefont {Kech}, \citenamefont
  {Schultz},\ and\ \citenamefont {Toigo}}]{carmeli2016stable}%
  \BibitemOpen
  \bibfield  {author} {\bibinfo {author} {\bibfnamefont {C.}~\bibnamefont
  {Carmeli}}, \bibinfo {author} {\bibfnamefont {T.}~\bibnamefont {Heinosaari}},
  \bibinfo {author} {\bibfnamefont {M.}~\bibnamefont {Kech}}, \bibinfo {author}
  {\bibfnamefont {J.}~\bibnamefont {Schultz}},\ and\ \bibinfo {author}
  {\bibfnamefont {A.}~\bibnamefont {Toigo}},\ }\bibfield  {title} {\bibinfo
  {title} {Stable pure state quantum tomography from five orthonormal bases},\
  }\href {https://doi.org/10.1209/0295-5075/115/30001} {\bibfield  {journal}
  {\bibinfo  {journal} {Europhys. Lett.}\ }\textbf {\bibinfo {volume} {115}},\
  \bibinfo {pages} {30001} (\bibinfo {year} {2016})}\BibitemShut {NoStop}%
\bibitem [{\citenamefont {Gross}\ \emph {et~al.}(2010)\citenamefont {Gross},
  \citenamefont {Liu}, \citenamefont {Flammia}, \citenamefont {Becker},\ and\
  \citenamefont {Eisert}}]{gross2010quantum}%
  \BibitemOpen
  \bibfield  {author} {\bibinfo {author} {\bibfnamefont {D.}~\bibnamefont
  {Gross}}, \bibinfo {author} {\bibfnamefont {Y.-K.}\ \bibnamefont {Liu}},
  \bibinfo {author} {\bibfnamefont {S.~T.}\ \bibnamefont {Flammia}}, \bibinfo
  {author} {\bibfnamefont {S.}~\bibnamefont {Becker}},\ and\ \bibinfo {author}
  {\bibfnamefont {J.}~\bibnamefont {Eisert}},\ }\bibfield  {title} {\bibinfo
  {title} {Quantum state tomography via compressed sensing},\ }\href
  {https://doi.org/10.1103/PhysRevLett.105.150401} {\bibfield  {journal}
  {\bibinfo  {journal} {Phys. Rev. Lett.}\ }\textbf {\bibinfo {volume} {105}},\
  \bibinfo {pages} {150401} (\bibinfo {year} {2010})}\BibitemShut {NoStop}%
\bibitem [{\citenamefont {Zheng}\ \emph {et~al.}(2015)\citenamefont {Zheng},
  \citenamefont {Yang}, \citenamefont {Liu}, \citenamefont {Liang},\ and\
  \citenamefont {Li}}]{softplus}%
  \BibitemOpen
  \bibfield  {author} {\bibinfo {author} {\bibfnamefont {H.}~\bibnamefont
  {Zheng}}, \bibinfo {author} {\bibfnamefont {Z.}~\bibnamefont {Yang}},
  \bibinfo {author} {\bibfnamefont {W.}~\bibnamefont {Liu}}, \bibinfo {author}
  {\bibfnamefont {J.}~\bibnamefont {Liang}},\ and\ \bibinfo {author}
  {\bibfnamefont {Y.}~\bibnamefont {Li}},\ }\bibfield  {title} {\bibinfo
  {title} {Improving deep neural networks using softplus units},\ }in\ \href
  {https://doi.org/10.1109/ijcnn.2015.7280459} {\emph {\bibinfo {booktitle}
  {2015 International Joint Conference on Neural Networks ({IJCNN})}}}\
  (\bibinfo  {publisher} {{IEEE}},\ \bibinfo {year} {2015})\BibitemShut
  {NoStop}%
\bibitem [{\citenamefont {Nair}\ and\ \citenamefont {Hinton}(2010)}]{relu}%
  \BibitemOpen
  \bibfield  {author} {\bibinfo {author} {\bibfnamefont {V.}~\bibnamefont
  {Nair}}\ and\ \bibinfo {author} {\bibfnamefont {G.~E.}\ \bibnamefont
  {Hinton}},\ }\bibfield  {title} {\bibinfo {title} {Rectified linear units
  improve restricted boltzmann machines},\ }in\ \href@noop {} {\emph {\bibinfo
  {booktitle} {Proceedings of the 27th International Conference on
  International Conference on Machine Learning}}},\ \bibinfo {series and
  number} {ICML'10}\ (\bibinfo  {publisher} {Omnipress},\ \bibinfo {address}
  {Madison, WI, USA},\ \bibinfo {year} {2010})\ p.\ \bibinfo {pages}
  {807–814}\BibitemShut {NoStop}%
\bibitem [{\citenamefont {Paszke}\ \emph {et~al.}(2019)\citenamefont {Paszke},
  \citenamefont {Gross}, \citenamefont {Massa}, \citenamefont {Lerer},
  \citenamefont {Bradbury}, \citenamefont {Chanan}, \citenamefont {Killeen},
  \citenamefont {Lin}, \citenamefont {Gimelshein}, \citenamefont {Antiga},
  \citenamefont {Desmaison}, \citenamefont {Kopf}, \citenamefont {Yang},
  \citenamefont {DeVito}, \citenamefont {Raison}, \citenamefont {Tejani},
  \citenamefont {Chilamkurthy}, \citenamefont {Steiner}, \citenamefont {Fang},
  \citenamefont {Bai},\ and\ \citenamefont {Chintala}}]{pytorch}%
  \BibitemOpen
  \bibfield  {author} {\bibinfo {author} {\bibfnamefont {A.}~\bibnamefont
  {Paszke}}, \bibinfo {author} {\bibfnamefont {S.}~\bibnamefont {Gross}},
  \bibinfo {author} {\bibfnamefont {F.}~\bibnamefont {Massa}}, \bibinfo
  {author} {\bibfnamefont {A.}~\bibnamefont {Lerer}}, \bibinfo {author}
  {\bibfnamefont {J.}~\bibnamefont {Bradbury}}, \bibinfo {author}
  {\bibfnamefont {G.}~\bibnamefont {Chanan}}, \bibinfo {author} {\bibfnamefont
  {T.}~\bibnamefont {Killeen}}, \bibinfo {author} {\bibfnamefont
  {Z.}~\bibnamefont {Lin}}, \bibinfo {author} {\bibfnamefont {N.}~\bibnamefont
  {Gimelshein}}, \bibinfo {author} {\bibfnamefont {L.}~\bibnamefont {Antiga}},
  \bibinfo {author} {\bibfnamefont {A.}~\bibnamefont {Desmaison}}, \bibinfo
  {author} {\bibfnamefont {A.}~\bibnamefont {Kopf}}, \bibinfo {author}
  {\bibfnamefont {E.}~\bibnamefont {Yang}}, \bibinfo {author} {\bibfnamefont
  {Z.}~\bibnamefont {DeVito}}, \bibinfo {author} {\bibfnamefont
  {M.}~\bibnamefont {Raison}}, \bibinfo {author} {\bibfnamefont
  {A.}~\bibnamefont {Tejani}}, \bibinfo {author} {\bibfnamefont
  {S.}~\bibnamefont {Chilamkurthy}}, \bibinfo {author} {\bibfnamefont
  {B.}~\bibnamefont {Steiner}}, \bibinfo {author} {\bibfnamefont
  {L.}~\bibnamefont {Fang}}, \bibinfo {author} {\bibfnamefont {J.}~\bibnamefont
  {Bai}},\ and\ \bibinfo {author} {\bibfnamefont {S.}~\bibnamefont
  {Chintala}},\ }\bibfield  {title} {\bibinfo {title} {{PyTorch: An Imperative
  Style, High-Performance Deep Learning Library}},\ }in\ \href
  {http://papers.neurips.cc/paper/9015-pytorch-an-imperative-style-high-performance-deep-learning-library.pdf}
  {\emph {\bibinfo {booktitle} {Advances in Neural Information Processing
  Systems 32}}},\ \bibinfo {editor} {edited by\ \bibinfo {editor}
  {\bibfnamefont {H.}~\bibnamefont {Wallach}}, \bibinfo {editor} {\bibfnamefont
  {H.}~\bibnamefont {Larochelle}}, \bibinfo {editor} {\bibfnamefont
  {A.}~\bibnamefont {Beygelzimer}}, \bibinfo {editor} {\bibfnamefont
  {F.}~\bibnamefont {d'Alché Buc}}, \bibinfo {editor} {\bibfnamefont
  {E.}~\bibnamefont {Fox}},\ and\ \bibinfo {editor} {\bibfnamefont
  {R.}~\bibnamefont {Garnett}}}\ (\bibinfo  {publisher} {Curran Associates,
  Inc.},\ \bibinfo {year} {2019})\ pp.\ \bibinfo {pages}
  {8024--8035}\BibitemShut {NoStop}%
\bibitem [{\citenamefont {Virtanen}\ \emph {et~al.}(2020)\citenamefont
  {Virtanen}, \citenamefont {Gommers}, \citenamefont {Oliphant}, \citenamefont
  {Haberland}, \citenamefont {Reddy}, \citenamefont {Cournapeau}, \citenamefont
  {Burovski}, \citenamefont {Peterson}, \citenamefont {Weckesser},
  \citenamefont {Bright}, \citenamefont {van~der Walt}, \citenamefont {Brett},
  \citenamefont {Wilson}, \citenamefont {Millman}, \citenamefont {Mayorov},
  \citenamefont {Nelson}, \citenamefont {Jones}, \citenamefont {Kern},
  \citenamefont {Larson}, \citenamefont {Carey}, \citenamefont {Polat},
  \citenamefont {Feng}, \citenamefont {Moore}, \citenamefont {VanderPlas},
  \citenamefont {Laxalde}, \citenamefont {Perktold}, \citenamefont {Cimrman},
  \citenamefont {Henriksen}, \citenamefont {Quintero}, \citenamefont {Harris},
  \citenamefont {Archibald}, \citenamefont {Ribeiro}, \citenamefont
  {Pedregosa}, \citenamefont {van Mulbregt}, \citenamefont {Vijaykumar},
  \citenamefont {Bardelli}, \citenamefont {Rothberg}, \citenamefont {Hilboll},
  \citenamefont {Kloeckner}, \citenamefont {Scopatz}, \citenamefont {Lee},
  \citenamefont {Rokem}, \citenamefont {Woods}, \citenamefont {Fulton},
  \citenamefont {Masson}, \citenamefont {Häggström}, \citenamefont
  {Fitzgerald}, \citenamefont {Nicholson}, \citenamefont {Hagen}, \citenamefont
  {Pasechnik}, \citenamefont {Olivetti}, \citenamefont {Martin}, \citenamefont
  {Wieser}, \citenamefont {Silva}, \citenamefont {Lenders}, \citenamefont
  {Wilhelm}, \citenamefont {Young}, \citenamefont {Price}, \citenamefont
  {Ingold}, \citenamefont {Allen}, \citenamefont {Lee}, \citenamefont {Audren},
  \citenamefont {Probst}, \citenamefont {Dietrich}, \citenamefont {Silterra},
  \citenamefont {Webber}, \citenamefont {Slavi{\v{c}}}, \citenamefont
  {Nothman}, \citenamefont {Buchner}, \citenamefont {Kulick}, \citenamefont
  {Schönberger}, \citenamefont {de~Miranda~Cardoso}, \citenamefont {Reimer},
  \citenamefont {Harrington}, \citenamefont {Rodr{\'{\i}}guez}, \citenamefont
  {Nunez-Iglesias}, \citenamefont {Kuczynski}, \citenamefont {Tritz},
  \citenamefont {Thoma}, \citenamefont {Newville}, \citenamefont {Kümmerer},
  \citenamefont {Bolingbroke}, \citenamefont {Tartre}, \citenamefont {Pak},
  \citenamefont {Smith}, \citenamefont {Nowaczyk}, \citenamefont {Shebanov},
  \citenamefont {Pavlyk}, \citenamefont {Brodtkorb}, \citenamefont {Lee},
  \citenamefont {McGibbon}, \citenamefont {Feldbauer}, \citenamefont {Lewis},
  \citenamefont {Tygier}, \citenamefont {Sievert}, \citenamefont {Vigna},
  \citenamefont {Peterson}, \citenamefont {More}, \citenamefont {Pudlik},
  \citenamefont {Oshima}, \citenamefont {Pingel}, \citenamefont {Robitaille},
  \citenamefont {Spura}, \citenamefont {Jones}, \citenamefont {Cera},
  \citenamefont {Leslie}, \citenamefont {Zito}, \citenamefont {Krauss},
  \citenamefont {Upadhyay}, \citenamefont {Halchenko},\ and\ \citenamefont
  {and}}]{2020SciPy-NMeth}%
  \BibitemOpen
  \bibfield  {author} {\bibinfo {author} {\bibfnamefont {P.}~\bibnamefont
  {Virtanen}}, \bibinfo {author} {\bibfnamefont {R.}~\bibnamefont {Gommers}},
  \bibinfo {author} {\bibfnamefont {T.~E.}\ \bibnamefont {Oliphant}}, \bibinfo
  {author} {\bibfnamefont {M.}~\bibnamefont {Haberland}}, \bibinfo {author}
  {\bibfnamefont {T.}~\bibnamefont {Reddy}}, \bibinfo {author} {\bibfnamefont
  {D.}~\bibnamefont {Cournapeau}}, \bibinfo {author} {\bibfnamefont
  {E.}~\bibnamefont {Burovski}}, \bibinfo {author} {\bibfnamefont
  {P.}~\bibnamefont {Peterson}}, \bibinfo {author} {\bibfnamefont
  {W.}~\bibnamefont {Weckesser}}, \bibinfo {author} {\bibfnamefont
  {J.}~\bibnamefont {Bright}}, \bibinfo {author} {\bibfnamefont {S.~J.}\
  \bibnamefont {van~der Walt}}, \bibinfo {author} {\bibfnamefont
  {M.}~\bibnamefont {Brett}}, \bibinfo {author} {\bibfnamefont
  {J.}~\bibnamefont {Wilson}}, \bibinfo {author} {\bibfnamefont {K.~J.}\
  \bibnamefont {Millman}}, \bibinfo {author} {\bibfnamefont {N.}~\bibnamefont
  {Mayorov}}, \bibinfo {author} {\bibfnamefont {A.~R.~J.}\ \bibnamefont
  {Nelson}}, \bibinfo {author} {\bibfnamefont {E.}~\bibnamefont {Jones}},
  \bibinfo {author} {\bibfnamefont {R.}~\bibnamefont {Kern}}, \bibinfo {author}
  {\bibfnamefont {E.}~\bibnamefont {Larson}}, \bibinfo {author} {\bibfnamefont
  {C.~J.}\ \bibnamefont {Carey}}, \bibinfo {author} {\bibfnamefont
  {{\.{I}}.}~\bibnamefont {Polat}}, \bibinfo {author} {\bibfnamefont
  {Y.}~\bibnamefont {Feng}}, \bibinfo {author} {\bibfnamefont {E.~W.}\
  \bibnamefont {Moore}}, \bibinfo {author} {\bibfnamefont {J.}~\bibnamefont
  {VanderPlas}}, \bibinfo {author} {\bibfnamefont {D.}~\bibnamefont {Laxalde}},
  \bibinfo {author} {\bibfnamefont {J.}~\bibnamefont {Perktold}}, \bibinfo
  {author} {\bibfnamefont {R.}~\bibnamefont {Cimrman}}, \bibinfo {author}
  {\bibfnamefont {I.}~\bibnamefont {Henriksen}}, \bibinfo {author}
  {\bibfnamefont {E.~A.}\ \bibnamefont {Quintero}}, \bibinfo {author}
  {\bibfnamefont {C.~R.}\ \bibnamefont {Harris}}, \bibinfo {author}
  {\bibfnamefont {A.~M.}\ \bibnamefont {Archibald}}, \bibinfo {author}
  {\bibfnamefont {A.~H.}\ \bibnamefont {Ribeiro}}, \bibinfo {author}
  {\bibfnamefont {F.}~\bibnamefont {Pedregosa}}, \bibinfo {author}
  {\bibfnamefont {P.}~\bibnamefont {van Mulbregt}}, \bibinfo {author}
  {\bibfnamefont {A.}~\bibnamefont {Vijaykumar}}, \bibinfo {author}
  {\bibfnamefont {A.~P.}\ \bibnamefont {Bardelli}}, \bibinfo {author}
  {\bibfnamefont {A.}~\bibnamefont {Rothberg}}, \bibinfo {author}
  {\bibfnamefont {A.}~\bibnamefont {Hilboll}}, \bibinfo {author} {\bibfnamefont
  {A.}~\bibnamefont {Kloeckner}}, \bibinfo {author} {\bibfnamefont
  {A.}~\bibnamefont {Scopatz}}, \bibinfo {author} {\bibfnamefont
  {A.}~\bibnamefont {Lee}}, \bibinfo {author} {\bibfnamefont {A.}~\bibnamefont
  {Rokem}}, \bibinfo {author} {\bibfnamefont {C.~N.}\ \bibnamefont {Woods}},
  \bibinfo {author} {\bibfnamefont {C.}~\bibnamefont {Fulton}}, \bibinfo
  {author} {\bibfnamefont {C.}~\bibnamefont {Masson}}, \bibinfo {author}
  {\bibfnamefont {C.}~\bibnamefont {Häggström}}, \bibinfo {author}
  {\bibfnamefont {C.}~\bibnamefont {Fitzgerald}}, \bibinfo {author}
  {\bibfnamefont {D.~A.}\ \bibnamefont {Nicholson}}, \bibinfo {author}
  {\bibfnamefont {D.~R.}\ \bibnamefont {Hagen}}, \bibinfo {author}
  {\bibfnamefont {D.~V.}\ \bibnamefont {Pasechnik}}, \bibinfo {author}
  {\bibfnamefont {E.}~\bibnamefont {Olivetti}}, \bibinfo {author}
  {\bibfnamefont {E.}~\bibnamefont {Martin}}, \bibinfo {author} {\bibfnamefont
  {E.}~\bibnamefont {Wieser}}, \bibinfo {author} {\bibfnamefont
  {F.}~\bibnamefont {Silva}}, \bibinfo {author} {\bibfnamefont
  {F.}~\bibnamefont {Lenders}}, \bibinfo {author} {\bibfnamefont
  {F.}~\bibnamefont {Wilhelm}}, \bibinfo {author} {\bibfnamefont
  {G.}~\bibnamefont {Young}}, \bibinfo {author} {\bibfnamefont {G.~A.}\
  \bibnamefont {Price}}, \bibinfo {author} {\bibfnamefont {G.-L.}\ \bibnamefont
  {Ingold}}, \bibinfo {author} {\bibfnamefont {G.~E.}\ \bibnamefont {Allen}},
  \bibinfo {author} {\bibfnamefont {G.~R.}\ \bibnamefont {Lee}}, \bibinfo
  {author} {\bibfnamefont {H.}~\bibnamefont {Audren}}, \bibinfo {author}
  {\bibfnamefont {I.}~\bibnamefont {Probst}}, \bibinfo {author} {\bibfnamefont
  {J.~P.}\ \bibnamefont {Dietrich}}, \bibinfo {author} {\bibfnamefont
  {J.}~\bibnamefont {Silterra}}, \bibinfo {author} {\bibfnamefont {J.~T.}\
  \bibnamefont {Webber}}, \bibinfo {author} {\bibfnamefont {J.}~\bibnamefont
  {Slavi{\v{c}}}}, \bibinfo {author} {\bibfnamefont {J.}~\bibnamefont
  {Nothman}}, \bibinfo {author} {\bibfnamefont {J.}~\bibnamefont {Buchner}},
  \bibinfo {author} {\bibfnamefont {J.}~\bibnamefont {Kulick}}, \bibinfo
  {author} {\bibfnamefont {J.~L.}\ \bibnamefont {Schönberger}}, \bibinfo
  {author} {\bibfnamefont {J.~V.}\ \bibnamefont {de~Miranda~Cardoso}}, \bibinfo
  {author} {\bibfnamefont {J.}~\bibnamefont {Reimer}}, \bibinfo {author}
  {\bibfnamefont {J.}~\bibnamefont {Harrington}}, \bibinfo {author}
  {\bibfnamefont {J.~L.~C.}\ \bibnamefont {Rodr{\'{\i}}guez}}, \bibinfo
  {author} {\bibfnamefont {J.}~\bibnamefont {Nunez-Iglesias}}, \bibinfo
  {author} {\bibfnamefont {J.}~\bibnamefont {Kuczynski}}, \bibinfo {author}
  {\bibfnamefont {K.}~\bibnamefont {Tritz}}, \bibinfo {author} {\bibfnamefont
  {M.}~\bibnamefont {Thoma}}, \bibinfo {author} {\bibfnamefont
  {M.}~\bibnamefont {Newville}}, \bibinfo {author} {\bibfnamefont
  {M.}~\bibnamefont {Kümmerer}}, \bibinfo {author} {\bibfnamefont
  {M.}~\bibnamefont {Bolingbroke}}, \bibinfo {author} {\bibfnamefont
  {M.}~\bibnamefont {Tartre}}, \bibinfo {author} {\bibfnamefont
  {M.}~\bibnamefont {Pak}}, \bibinfo {author} {\bibfnamefont {N.~J.}\
  \bibnamefont {Smith}}, \bibinfo {author} {\bibfnamefont {N.}~\bibnamefont
  {Nowaczyk}}, \bibinfo {author} {\bibfnamefont {N.}~\bibnamefont {Shebanov}},
  \bibinfo {author} {\bibfnamefont {O.}~\bibnamefont {Pavlyk}}, \bibinfo
  {author} {\bibfnamefont {P.~A.}\ \bibnamefont {Brodtkorb}}, \bibinfo {author}
  {\bibfnamefont {P.}~\bibnamefont {Lee}}, \bibinfo {author} {\bibfnamefont
  {R.~T.}\ \bibnamefont {McGibbon}}, \bibinfo {author} {\bibfnamefont
  {R.}~\bibnamefont {Feldbauer}}, \bibinfo {author} {\bibfnamefont
  {S.}~\bibnamefont {Lewis}}, \bibinfo {author} {\bibfnamefont
  {S.}~\bibnamefont {Tygier}}, \bibinfo {author} {\bibfnamefont
  {S.}~\bibnamefont {Sievert}}, \bibinfo {author} {\bibfnamefont
  {S.}~\bibnamefont {Vigna}}, \bibinfo {author} {\bibfnamefont
  {S.}~\bibnamefont {Peterson}}, \bibinfo {author} {\bibfnamefont
  {S.}~\bibnamefont {More}}, \bibinfo {author} {\bibfnamefont {T.}~\bibnamefont
  {Pudlik}}, \bibinfo {author} {\bibfnamefont {T.}~\bibnamefont {Oshima}},
  \bibinfo {author} {\bibfnamefont {T.~J.}\ \bibnamefont {Pingel}}, \bibinfo
  {author} {\bibfnamefont {T.~P.}\ \bibnamefont {Robitaille}}, \bibinfo
  {author} {\bibfnamefont {T.}~\bibnamefont {Spura}}, \bibinfo {author}
  {\bibfnamefont {T.~R.}\ \bibnamefont {Jones}}, \bibinfo {author}
  {\bibfnamefont {T.}~\bibnamefont {Cera}}, \bibinfo {author} {\bibfnamefont
  {T.}~\bibnamefont {Leslie}}, \bibinfo {author} {\bibfnamefont
  {T.}~\bibnamefont {Zito}}, \bibinfo {author} {\bibfnamefont {T.}~\bibnamefont
  {Krauss}}, \bibinfo {author} {\bibfnamefont {U.}~\bibnamefont {Upadhyay}},
  \bibinfo {author} {\bibfnamefont {Y.~O.}\ \bibnamefont {Halchenko}},\ and\
  \bibinfo {author} {\bibfnamefont {Y.~V.-B.}\ \bibnamefont {and}},\ }\bibfield
   {title} {\bibinfo {title} {{SciPy} 1.0: fundamental algorithms for
  scientific computing in python},\ }\href
  {https://doi.org/10.1038/s41592-019-0686-2} {\bibfield  {journal} {\bibinfo
  {journal} {Nat. Methods}\ }\textbf {\bibinfo {volume} {17}},\ \bibinfo
  {pages} {261} (\bibinfo {year} {2020})}\BibitemShut {NoStop}%
\bibitem [{\citenamefont {Koenig}\ and\ \citenamefont
  {Smolin}(2014)}]{Koenig_2014}%
  \BibitemOpen
  \bibfield  {author} {\bibinfo {author} {\bibfnamefont {R.}~\bibnamefont
  {Koenig}}\ and\ \bibinfo {author} {\bibfnamefont {J.~A.}\ \bibnamefont
  {Smolin}},\ }\bibfield  {title} {\bibinfo {title} {How to efficiently select
  an arbitrary clifford group element},\ }\bibfield  {journal} {\bibinfo
  {journal} {Journal of Mathematical Physics}\ }\textbf {\bibinfo {volume}
  {55}},\ \href {https://doi.org/10.1063/1.4903507} {10.1063/1.4903507}
  (\bibinfo {year} {2014})\BibitemShut {NoStop}%
\bibitem [{\citenamefont {Biswas}\ \emph {et~al.}(2021)\citenamefont {Biswas},
  \citenamefont {Biswas},\ and\ \citenamefont {Sen}}]{haar-measure}%
  \BibitemOpen
  \bibfield  {author} {\bibinfo {author} {\bibfnamefont {G.}~\bibnamefont
  {Biswas}}, \bibinfo {author} {\bibfnamefont {A.}~\bibnamefont {Biswas}},\
  and\ \bibinfo {author} {\bibfnamefont {U.}~\bibnamefont {Sen}},\ }\bibfield
  {title} {\bibinfo {title} {Inhibition of spread of typical bipartite and
  genuine multiparty entanglement in response to disorder},\ }\href
  {https://doi.org/10.1088/1367-2630/ac37c8} {\bibfield  {journal} {\bibinfo
  {journal} {New J. Phys.}\ }\textbf {\bibinfo {volume} {23}},\ \bibinfo
  {pages} {113042} (\bibinfo {year} {2021})}\BibitemShut {NoStop}%
\bibitem [{\citenamefont {Diamond}\ and\ \citenamefont
  {Boyd}(2016)}]{diamond2016cvxpy}%
  \BibitemOpen
  \bibfield  {author} {\bibinfo {author} {\bibfnamefont {S.}~\bibnamefont
  {Diamond}}\ and\ \bibinfo {author} {\bibfnamefont {S.}~\bibnamefont {Boyd}},\
  }\bibfield  {title} {\bibinfo {title} {Cvxpy: A python-embedded modeling
  language for convex optimization},\ }\href@noop {} {\bibfield  {journal}
  {\bibinfo  {journal} {J. Mach. Learn. Res.}\ }\textbf {\bibinfo {volume}
  {17}},\ \bibinfo {pages} {2909–2913} (\bibinfo {year} {2016})}\BibitemShut
  {NoStop}%
\bibitem [{\citenamefont {Li}\ \emph {et~al.}(2017)\citenamefont {Li},
  \citenamefont {Huang}, \citenamefont {Luo}, \citenamefont {Li}, \citenamefont
  {Lu},\ and\ \citenamefont {Zeng}}]{li2017optimal}%
  \BibitemOpen
  \bibfield  {author} {\bibinfo {author} {\bibfnamefont {J.}~\bibnamefont
  {Li}}, \bibinfo {author} {\bibfnamefont {S.}~\bibnamefont {Huang}}, \bibinfo
  {author} {\bibfnamefont {Z.}~\bibnamefont {Luo}}, \bibinfo {author}
  {\bibfnamefont {K.}~\bibnamefont {Li}}, \bibinfo {author} {\bibfnamefont
  {D.}~\bibnamefont {Lu}},\ and\ \bibinfo {author} {\bibfnamefont
  {B.}~\bibnamefont {Zeng}},\ }\bibfield  {title} {\bibinfo {title} {Optimal
  design of measurement settings for quantum-state-tomography experiments},\
  }\href {https://doi.org/10.1103/PhysRevA.96.032307} {\bibfield  {journal}
  {\bibinfo  {journal} {Phys. Rev. A}\ }\textbf {\bibinfo {volume} {96}},\
  \bibinfo {pages} {032307} (\bibinfo {year} {2017})}\BibitemShut {NoStop}%
\bibitem [{\citenamefont {Gühne}\ and\ \citenamefont
  {T{\'{o}}th}(2009)}]{G_hne_2009}%
  \BibitemOpen
  \bibfield  {author} {\bibinfo {author} {\bibfnamefont {O.}~\bibnamefont
  {Gühne}}\ and\ \bibinfo {author} {\bibfnamefont {G.}~\bibnamefont
  {T{\'{o}}th}},\ }\bibfield  {title} {\bibinfo {title} {Entanglement
  detection},\ }\href {https://doi.org/10.1016/j.physrep.2009.02.004}
  {\bibfield  {journal} {\bibinfo  {journal} {Phys. Rep.}\ }\textbf {\bibinfo
  {volume} {474}},\ \bibinfo {pages} {1} (\bibinfo {year} {2009})}\BibitemShut
  {NoStop}%
\bibitem [{\citenamefont {Duan}(2009)}]{duan2009super}%
  \BibitemOpen
  \bibfield  {author} {\bibinfo {author} {\bibfnamefont {R.}~\bibnamefont
  {Duan}},\ }\bibfield  {title} {\bibinfo {title} {Super-activation of
  zero-error capacity of noisy quantum channels},\ }\href@noop {} {\bibfield
  {journal} {\bibinfo  {journal} {arXiv preprint arXiv:0906.2527}\ } (\bibinfo
  {year} {2009})}\BibitemShut {NoStop}%
\bibitem [{\citenamefont {Johnston}\ \emph {et~al.}(2022)\citenamefont
  {Johnston}, \citenamefont {Lovitz},\ and\ \citenamefont
  {Vijayaraghavan}}]{johnston2022complete}%
  \BibitemOpen
  \bibfield  {author} {\bibinfo {author} {\bibfnamefont {N.}~\bibnamefont
  {Johnston}}, \bibinfo {author} {\bibfnamefont {B.}~\bibnamefont {Lovitz}},\
  and\ \bibinfo {author} {\bibfnamefont {A.}~\bibnamefont {Vijayaraghavan}},\
  }\bibfield  {title} {\bibinfo {title} {Complete hierarchy of linear systems
  for certifying quantum entanglement of subspaces},\ }\href
  {https://doi.org/10.1103/PhysRevA.106.062443} {\bibfield  {journal} {\bibinfo
   {journal} {Phys. Rev. A}\ }\textbf {\bibinfo {volume} {106}},\ \bibinfo
  {pages} {062443} (\bibinfo {year} {2022})}\BibitemShut {NoStop}%
\bibitem [{\citenamefont {Zhang}\ and\ \citenamefont
  {Zhu}(2023)}]{github-repo}%
  \BibitemOpen
  \bibfield  {author} {\bibinfo {author} {\bibfnamefont {C.}~\bibnamefont
  {Zhang}}\ and\ \bibinfo {author} {\bibfnamefont {X.}~\bibnamefont {Zhu}},\
  }\href {https://github.com/husisy/udap-public} {\bibinfo {title} {{UDA-UDP
  GitHub repository}}} (\bibinfo {year} {2023})\BibitemShut {NoStop}%
\end{thebibliography}%
\end{document}